\documentclass[a4paper,10pt,times,twocolumn]{article}

\usepackage{moreverb,url}
\usepackage[colorlinks,bookmarksopen,bookmarksnumbered,citecolor=red,urlcolor=red]{hyperref}
\usepackage{tabularx}
\usepackage{multirow}
\usepackage{amssymb}
\usepackage{booktabs}
\usepackage{color,soul}
\usepackage{subfigure}
\usepackage{graphicx}
\usepackage{gensymb}
\usepackage[export]{adjustbox}
\usepackage{wrapfig}
\usepackage{float}
\usepackage{comment}
\usepackage{textcomp}
\usepackage[switch]{lineno} 
\usepackage{lipsum}
\usepackage{setspace}
\usepackage[normalem]{ulem}
\usepackage{amsmath}

\newcommand\BibTeX{{\rmfamily B\kern-.05em \textsc{i\kern-.025em b}\kern-.08em
T\kern-.1667em\lower.7ex\hbox{E}\kern-.125emX}}

\date{}

\begin{document}

\title{MPCA-based Domain Adaptation for Transfer Learning with Ultrasonic Guided Waves}

\author{Lucio Pinello, Francesco Cadini and Luca Lomazzi}

\maketitle

This work is licensed under a CC BY 4.0 license.

\begin{abstract}

Ultrasonic Guided Waves (UGWs) represent a promising diagnostic tool for Structural Health Monitoring (SHM) in thin-walled structures, and their integration with machine learning (ML) algorithms is increasingly being adopted to enable real-time monitoring capabilities.
However, the large-scale deployment of UGW-based ML methods is constrained by data scarcity and limited generalisation across different materials and sensor configurations.
To address these limitations, this work proposes a novel transfer learning (TL) framework based on Multilinear Principal Component Analysis (MPCA).
First, a Convolutional Neural Network (CNN) for regression is trained to perform damage localisation for a plated structure. Then, MPCA and fine-tuning are combined to have the CNN work for a different plate. By jointly applying MPCA to the source and target domains, the method extracts shared latent features, enabling effective domain adaptation without requiring prior assumptions about dimensionality. Following MPCA, fine-tuning enables adapting the pre-trained CNN to a new domain without the need for a large training dataset.
The proposed MPCA-based TL method was tested against 12 case studies involving different composite materials and sensor arrays.
Statistical metrics were used to assess domains alignment both before and after MPCA, and the results demonstrate a substantial reduction in localisation error compared to standard TL techniques. Hence, the proposed approach emerges as a robust, data-efficient, and statistically based TL framework for UGW-based SHM.

\end{abstract}
\vspace{1.5cm}
Keywords: UGW, SHM, transfer learning, domain adaptation, damage localisation, CNN\\

\section*{Acronyms}
\begin{table}[ht!]
    \begin{tabular}{l l}
    \textbf{UGW} & \begin{tabular}{l}Ultrasonic Guided Wave\end{tabular}\\
     & \\
    \textbf{CNN} & \begin{tabular}{l}Convolutional Neural Network\end{tabular}\\
     & \\
    \textbf{MPCA} & \begin{tabular}{l} Multilinear Principal Component \\ Analysis \end{tabular}\\
     & \\
    \textbf{S-CNN} & \begin{tabular}{l} Damage localisation obtained \\ with CNNs trained on the source \\ domain\end{tabular}\\
     & \\
    \textbf{T-CNN} & \begin{tabular}{l} Damage localisation obtained \\ with CNNs trained on the target \\ domain\end{tabular}\\
    \end{tabular}
\end{table}
\begin{table}[ht!]
    \begin{tabular}{l l}
    \textbf{FT} & \begin{tabular}{l} Damage localisation \\ obtained with CNNs \\ trained on the source \\ domain and fine-tuned \\ on the target domain\end{tabular}\\
     & \\
    \textbf{MPCA-FT} & \begin{tabular}{l} Damage localisation \\ obtained with  CNNs \\ trained on the mapped \\ source domain and \\ fine-tuned on the \\ mapped target domain\end{tabular}\\
     & \\
    \textbf{KL} & \begin{tabular}{l}Kullback-Leibler \\ Divergence\end{tabular}\\
     & \\
    \textbf{JSD} & \begin{tabular}{l}Jensen–Shannon \\ Divergence\end{tabular}\\
     & \\
     \textbf{$\chi^2$} & \begin{tabular}{l}Chi-Squared \\ Distance\end{tabular}\\
     & \\
     \textbf{B} & \begin{tabular}{l}Bhattacharyya  \\ Distance\end{tabular}\\
     & \\
    \textbf{EMD} & \begin{tabular}{l}Earth's Mover \\ Distance\end{tabular}\\
    \end{tabular}
\end{table}

\section{Introduction}\label{sec:intro}
Structural Health Monitoring (SHM) is gaining popularity as a way to ensure the safe operational life of engineering structures.
This is primarily driven by the need to avoid failures, especially of critical components, and to transition from standard time-scheduled monitoring and maintenance approaches to real-time and condition-based strategies \cite{CBM1,CBM2,HMAS}.\\
Within this framework, Ultrasonic Guided Waves (UGWs), also known as Lamb waves, are commonly used to asses the health state of thin-walled structures.
UGWs can travel long distances with low attenuation, allowing for a relatively low number of sensors even in large structures \cite{UGW1}.
This is a significant advantage compared to bulk waves, for which non-destructive evaluation (NDE) is performed by means of time-consuming point-to-point methods\cite{UGW2}.
In addition, UGWs are sensitive to small-scale damage, as the smallest detectable damage is on the order of the signal wavelength \cite{bookUGW}.
Another advantage of using UGWs is the ability to monitor a variety of different damage types, such as corrosion, crack, delamination, debonding, and others \cite{damUGW1,damUGW2,damUGW3,damUGW4}.
However, conventional damage diagnosis methods rely on expert knowledge, and require manual tuning of processing parameters and feature extraction to compute damage indices. Furthermore, the optimal subjective parameters required to achieve satisfactory damage diagnosis performance are specific to each application.\\
To overcome the limitations mentioned above, machine learning (ML)-based methods have gained popularity, as they enable automation and enhance both feature extraction and the computation of damage indices \cite{ML1,ML2,ML3,ML4,ML5}.
Among ML techniques, particular attention has been paid to Convolutional Neural Networks (CNNs), due to their ability to extract damage-sensitive features from high-dimensional raw data\cite{CNN1,CNN2,CNN3,CNN4,CNN5,CNN6,CNN7,CNN8,CNN9}.
Zhang et al. \cite{CNN1} used a one-dimensional CNN to perform damage localisation on aluminium plates based on a time-varying damage index. However, damage indices were still manually computed.
Similarly, Shattarifar et al. \cite{CNN2} used the downsampled variance of the cross-correlated aggregated measurements (DVCAM) pre-processing method, followed by a data augmentation step, to classify damage using a CNN.
In contrast, Rai et al. \cite{CNN3} developed a CNN for classification that takes in raw UGW data from experiments and simulations.
Similarly, Sawant et al. \cite{CNN4} used a CNN to detect damage, again using raw data as input, and then used a signal difference coefficient (SDC) algorithm to localise damage.
An intermediate processing step was performed to remove temperature effects on the CNN output.
Schackmann et al. \cite{CNN5} used a similar approach for damage localisation.
The raw UGW data underwent a Gramian angular field (GAF) transformation to obtain 2D images for the CNN.
Then, the CNN was trained to detect damage for each sensor pair, and the Receiver Operating Characteristic (ROC) curve was computed for each pair of sensors. The area under those curves was used as a metric to obtain a damage probability map of the structure.
A step forward was made by Lomazzi et al. \cite{LOMAZZI2023106003} and Gonzalez-Jimenez et al. \cite{CNN9}, who developed CNNs for regression to localise and quantify damage in metal and composite plates, respectively.
However, all ML methods presented above share the same limitation: a massive amount of data is required to train the networks. This limitation makes it impossible to deploy such methods in real-world, large-scale scenarios due to data scarcity.\\
Transfer Learning (TL) techniques can be applied to overcome this limitation.
TL focuses on maximising the knowledge acquired in a given domain, called the source domain, to improve knowledge in a new domain, called the target domain, for which data is limited \cite{TLsurvey,SVMclass,TLtheory}.
It may allow acquiring a limited amount of experimental data from a real-world scenario, and leverage knowledge from other sources, such as data obtained from different structures or from numerical simulations, to support the learning process of ML models.
Additionally, TL also addresses those scenarios in which tasks, and not only domains, are different \cite{TLsurvey}.
According to Ref.~\cite{TLsurvey}, if labelled data are available for both the source and target domains, and the target task differs from the source task, then TL is termed inductive TL.
Instead, transductive TL, or Domain Adaptation (DA), is applied if labelled data are available only for the source domain, but the source task is the same as the target task.
Finally, unsupervised TL is used when no labelled data is available, and the target task is different, but related, to the source one.
The classification presented above is based on tasks and domains, that is, the objective of the learning process and the methods used to learn.
However, a classification of the TL methods can also be made based on knowledge transfer between domains \cite{TLsurvey,TL_DR}.
Instance-based TL assumes that some data from the source domain can enhance the learning process on the target domain, while others may be detrimental.
Thus, techniques such as importance sampling or reweighting are employed to improve training by increasing the contribution of source domain data that have a beneficial effect on model performance.
In contrast, feature-representation TL seeks a latent representation where the source and target domain distributions are well-aligned. 
Within this latter category, Transfer Component Analysis (TCA) methods are commonly employed to identify a transformation that aligns the latent spaces of the domains.
Finally, parameter-based TL assumes that there are common hyperparameters between the source task and the target task, so the goal is to find the shared parameters required to transfer knowledge.\\
Within the UGW  field, TL has been implemented to address the data scarcity problem \cite{Sim2Real}, the presence of vibrations with different amplitudes in real operations \cite{multipleSources}, to minimise the number of training parameters \cite{caeTL}, and to adapt the knowledge acquired on a structure to a new one \cite{fullFT,convLSTM}.
Alguri et al. \cite{Sim2Real} used data from simulations of UGW propagation in aluminium plates to train a dictionary learning algorithm, and then applied it to data from real aluminium plates.
Although dictionary learning does not strictly fall into any transfer learning class, the way it was used in the study resembled a transductive TL application.
Yang et al. \cite{multipleSources} employed domain adaptation to align the distributions of UGW data acquired under vibrations of varying amplitudes.
The proposed method demonstrated good performance, although it was limited to vibrations excited at a single frequency.
Sawant et al. \cite{caeTL} employed a parameter-based TL framework on a composite plate with a rectangular piezoelectric (PZT) sensor network.
They trained an autoencoder to reconstruct the UGW signal from a pair of PZT sensors.
Then, assuming that the operations performed by the encoder are independent of the specific sensor pair, they trained a network per pair by fine-tuning the decoder only.
Finally, SDC was used to perform damage localisation based on the reconstruction error of the autoencoders, with good results.
Similarly, Rai et al. \cite{fullFT} trained a 1D convolutional autoencoder on healthy UGW data.
The encoder part of the autoencoder was then separated and used in combination with a CNN to perform damage localisation on a new dataset.
After that, both the pre-trained autoencoder and the added CNN were re-trained on the target domain.
Finally, Zhang et al. \cite{convLSTM} employed a joint domain adaptation technique to adapt the conditional distribution and the marginal distribution of the source domains (both healthy and damaged conditions) and the target domain (healthy condition).
Then, a convolutional Long Short-Term Memory (ConvLSTM) network was employed to learn the mapping function between the adapted training samples and the damage indices.
They successfully transferred knowledge between an aluminium plate and a composite plate, both with only two PZTs, and a composite plate with a circular sensor array.
Damage indices were used to obtain a probability map to localise damage.
However, the performance of the method depends on the choice of the convolutional kernel size, and damage index thresholding is needed for effective damage localisation.
Furthermore, this approach requires that the user manually tunes the process parameters to achieve good performance.\\
Inspired by the approaches described above, this work proposes a framework employing CNNs for damage localisation across varying combinations of materials and sensor networks.
The TL method employed follows a two-step procedure.
First, Multilinear Principal Component Analysis (MPCA) is applied to the combined source and target domains.
Then, fine-tuning is performed to have the network trained on the source domain work in the target domain.
This technique falls under the DA category, as it addresses distribution mismatches between the source and target domains while aiming to accomplish the same task, i.e., damage localisation.
However, when viewed from the perspective of the knowledge transfer between domains, this work lies at the intersection of feature-based and parameter-based approaches, corresponding to the use of MPCA and fine-tuning, respectively.
In this work, MPCA was chosen over other methods because of its superior capability to generate efficient latent representations.
MPCA was applied to a single database consisting of both the source and the target domain data.
Combining source and target domains in a single tensor makes MPCA identify the principal components shared by the two domains, and compress them into a common latent space.
Then, data are reconstructed by projecting the original tensor through the selected principal components, obtaining a lower-dimensionality dataset.
This approach is different from the traditional one, which leverages feature extraction to obtain separate latent representations of the source and target domains, and TCA techniques to build a transformation that maps the latent representations into a reduced Kernel Hilbert space \cite{TCA,TCA_MSSP}.
To the best of the authors' knowledge, this approach has not been proposed in the field of UGW-based SHM yet, where MPCA has only been used for dimensionality reduction.
A preliminary version of this approach has already been presented by the authors in Ref.~\cite{EWSHM24-TL}.
The main advantages of this approach over other feature-based domain adaptation techniques are: (i) the dimensionality of the latent (or feature) space is not chosen a priori but is instead determined by the amount of variability to retain, and (ii) principal components are computed using well-established MPCA equations, rather than through black-box mechanisms.
Additionally, the proposed algorithm is flexible and does not require adaptation for different applications, except for the selection of the new percentage of variance to retain.\\
The proposed framework was tested against the database presented in Ref.~\cite{CNN9}. Particularly, TL was applied on three composite plates made of different materials - $G16$, $K8$, and $K2G4S$ - onto which a circular and a rectangular sensor network were installed.
Two tasks were performed: (i) TL was applied to handle plates made of different materials but sharing the same sensor array, and (ii) TL was used to perform damage localisation on plates made of the same material, but with different sensor network layouts.
The performance of the proposed framework was tested under data scarcity by halving the data available for the target domain.
This reduced dataset was shown to be insufficient for training the CNN for regression on the target domain alone.
The data set generated and analysed during the current study is available in the Zenodo repository named UGW-3Mat-2SN \cite{UGW3MAT2SN}. 
The manuscript is organised as follows. Section \ref{sec:MM} describes the database, the CNN architecture, and the proposed MPCA-based transfer learning approach.
Section \ref{sec:caseStudy} presents the results obtained from the 12 case studies.
Finally, Section \ref{sec:conc} summarises the conclusions and outlines directions for future work.

\section{Materials and Methods}\label{sec:MM}
%
%
%
%

\subsection{Database}\label{sec:mat}
Three $200 $mm$ \times 300 $mm$ \times 4$ $mm$ panels made of different $0^\circ/90^\circ$ woven fabrics were manufactured.
The fabric materials employed were Kevlar$^{\circledR}$ 29 from Dupont$^{\circledR}$ \cite{dupont2019kevlar}, and 8 Harness Satin S2-Glass from Hexcel$^{\circledR}$ \cite{hexcel2012handbok}.
The matrix was epoxy resin AR260 \cite{resin} with AH260 hardener from Barracuda Advanced Composites \cite{hardener}.
The nomenclature and layout information of the samples are presented in Table \ref{tab:fabric}.

\begin{table}[htbp]
    \centering
    \caption{Nomenclature and composition of the composite plates used in the experimental campaign.}
    \label{tab:fabric}
    \adjustbox{width=\columnwidth}{
    \begin{tabular}{l|cc}
        Nomenclature & Fabric Material & Number of Layers \\
        \hline
        $K8$ & Kevlar 29 & 8 \\
        $G16$ & S2-Glass fibre & 16 \\
        $K2G4S$ & \begin{tabular}{c} Kevlar 29 and \\ S2-Glass \end{tabular} & \begin{tabular}{c} 4 Kevlar (K) and 8 S2-Glass \\ with layup $\big[K_2G_4\big]_S$ \end{tabular}
    \end{tabular}
    }
\end{table}

Two sensor arrays — a circular and a rectangular array — of eight PZT transducers each were installed on each plate.
Both sensor networks were centred on the centre of the panel.
The transducers were 1 $mm$ thick disks with a diameter of 5 $mm$ and electrodes wrapped around the edges, made of PIC255 \cite{pi2017piezo} and manufactured by PI Ceramic GmbH.
The rectangular array covered a scanning area of $160 mm \times 240 mm$, whereas the circular array allowed studying damage within a circular area with a diameter of 160 $mm$.
Table \ref{tab:pztLoc} reports the location of the PZTs installed on the plates.
PZTs from 1 to 9 constitute the circular sensor array.
Instead, the rectangular sensor network is formed by PZT 3, PZT 7, and PZTs from 9 to 14.
Thus, PZT3 and PZT7 are common to both sensor arrays.

\begin{table}[ht]
\centering
\caption{Location and identifier of PZT sensors on the plates.}
\label{tab:pztLoc}
\adjustbox{width=\columnwidth}{
\begin{tabular}{l|cc c l|cc}
Sensor ID & x $[mm]$ & y $[mm]$ & & Sensor ID & x $[mm]$ & y $[mm]$ \\
\cline{1-3}  \cline{5-7}
PZT1 & 100 & 230 & & PZT8 & 157 & 207 \\
PZT2 & 43  & 207 & & PZT9 & 20  & 30 \\
PZT3 & 20  & 150 & & PZT10 & 20  & 270 \\
PZT4 & 43  & 94 & & PZT11 & 100 & 270 \\
PZT5 & 100 & 70 & & PZT12 & 180 & 270 \\
PZT6 & 157 & 94 & & PZT13 & 180 & 30 \\
PZT7 & 180 & 150 & & PZT14 & 100 & 30 \\
\end{tabular}
}
\end{table}

\begin{figure}[hbtp]
    \centering
    \includegraphics[width=\columnwidth]{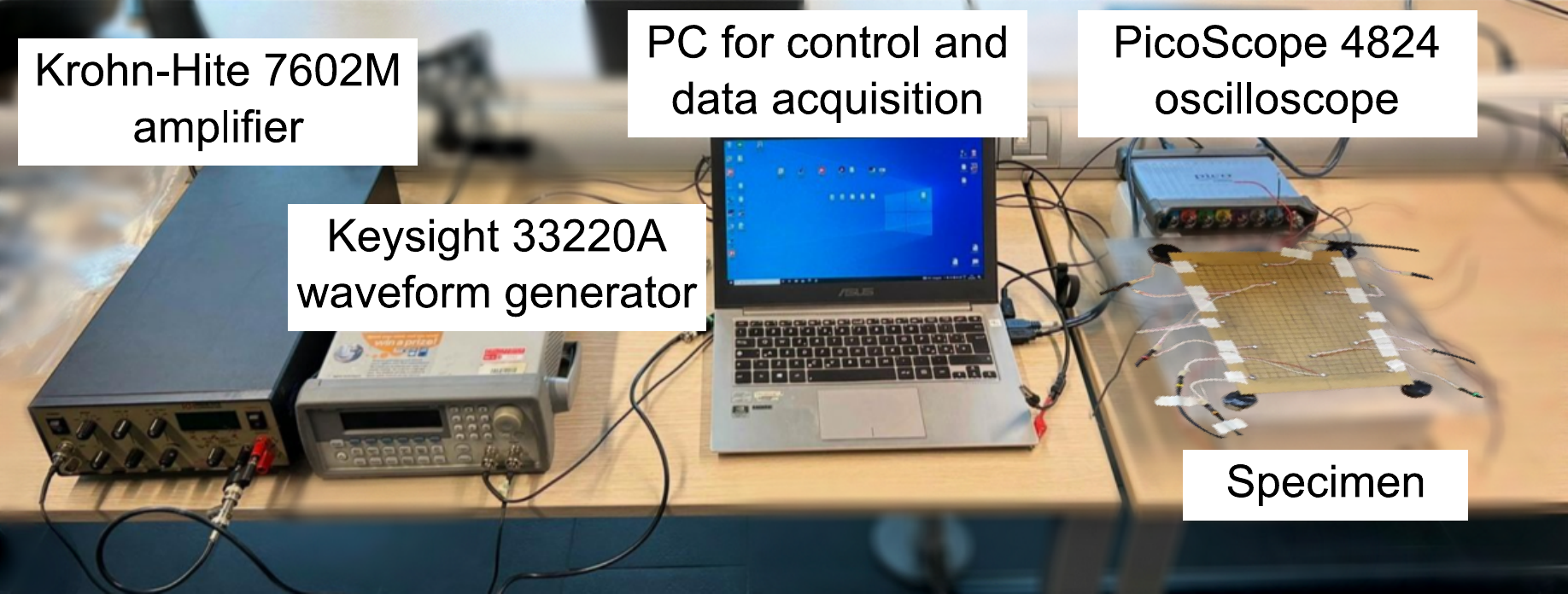}
    \caption{Experimental setup (from the left to the right): Krohn-Hite 7602M amplifier, Keysight 33220A waveform generator, PC for control and data acquisition, PicoScope 4824 oscilloscope, and a representative specimen.}
    \label{fig:expsetup}
\end{figure}

UGWs were excited using a Keysight Technologies$^{\circledR}$ model 33220A waveform generator, outputting a 10 $V$ excitation signal, in series with a Krohn-Hite$^{\circledR}$ corporation model 7602M amplifier, which amplified the signal up to 120 $V_{pp}$ to drive the PZT disks.
The excitation signal was a 150 $kHz$ five-cycle tone burst modulated with a Hann window.
The selected frequency was chosen after analysing the dispersion curves as a trade-off to allow for identifying minor damage while preventing high-order modes.
UGWs were acquired with a 4 $MHz$ sampling frequency using a PicoScope$^{\circledR}$ 4824A oscilloscope with eight available channels.
Figure \ref{fig:expsetup} shows the described experimental setup involved in the investigation.
To remove noise, acquired signals were filtered by a band-pass Butterworth filter with lower and upper cut-off frequencies of 50 and 250 $kHz$, respectively.
UGWs were excited and acquired using the pitch–catch approach.
That is, one PZT at a time was used to excite the wave, whereas the rest of the transducers served as sensors.
Each wave was acquired 20 times, and the final signal was obtained by calculating the median value of all the acquisitions.

The pseudo-damage approach was employed to simulate damage, as it enables a large number of tests with damage in multiple locations using a single specimen, thereby making experimental campaigns more cost-effective.
A 20 $mm$ diameter vinyl rubber tape (Scotch Vinyl Mastic Rolls 2210 \cite{scotchVinyl}), commonly used for isolation and vibration damping, was used to simulate damage.
Specifically, a circular tape was placed on the surface of the panels at a time, considering a total of 32 different positions, as reported in the Table \ref{tab:damLoc}.

\begin{table}[ht]
\centering
\caption{Location and identifier of simulated damages on the plates.}
\label{tab:damLoc}
\adjustbox{width=\columnwidth}{
\begin{tabular}{l|cc c l|cc}
Damage ID & x $[mm]$ & y $[mm]$ & & Damage ID & x $[mm]$ & y $[mm]$ \\
\cline{1-3}  \cline{5-7}
D1  & 35  & 255 & & D17 & 35  & 135 \\
D2  & 65  & 255 & & D18 & 65  & 135 \\
D3  & 125 & 255 & & D19 & 95  & 135 \\
D4  & 155 & 255 & & D20 & 125 & 135 \\
D5  & 35  & 225 & & D21 & 155 & 135 \\
D6  & 65  & 225 & & D22 & 65  & 105 \\
D7  & 125 & 225 & & D23 & 95  & 105 \\
D8  & 155 & 225 & & D24 & 125 & 105 \\
D9  & 65  & 195 & & D25 & 35  & 75  \\
D10 & 95  & 195 & & D26 & 65  & 75  \\
D11 & 125 & 195 & & D27 & 125 & 75  \\
D12 & 35  & 165 & & D28 & 155 & 75  \\
D13 & 65  & 165 & & D29 & 35  & 45  \\
D14 & 95  & 165 & & D30 & 65  & 45  \\
D15 & 125 & 165 & & D31 & 125 & 45  \\
D16 & 155 & 165 & & D32 & 155 & 45  \\
\end{tabular}
}
\end{table}

\begin{figure}[hbtp]
    \centering
    \subfigure[]{\includegraphics[width=0.75\columnwidth]{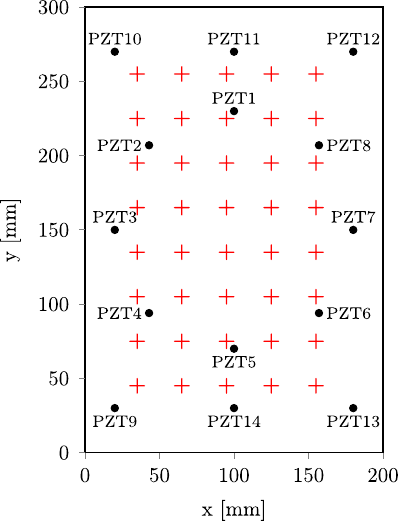}}
    \subfigure[]{\includegraphics[width=0.75\columnwidth]{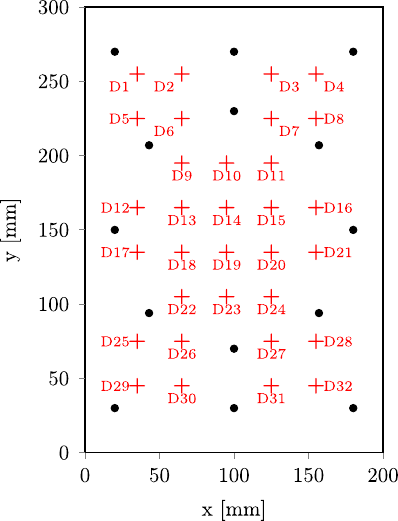}}
    \caption{Sensor (black dots) and damage (red crosses) locations on the plate with (a) PZT IDs and (b) damage IDs.}
    \label{fig:loc}
\end{figure}

The positions of the sensors and the damage locations are shown in Figures \ref{fig:loc}(a) and \ref{fig:loc}(b), respectively.
All damage locations, e.g., 32 damages, were taken into account for the rectangular array.
In contrast, only damage positions from D9 to D24 were considered for the circular array, for a total of 16 damage locations.

Following the approach described above, five different datasets were created. That is, a dataset per material type and sensor array was created, except for the rectangular sensor network of the $K8$ plate, which failed during testing.

\begin{figure}[hbtp]
    \centering
    \includegraphics[width=0.75\columnwidth]{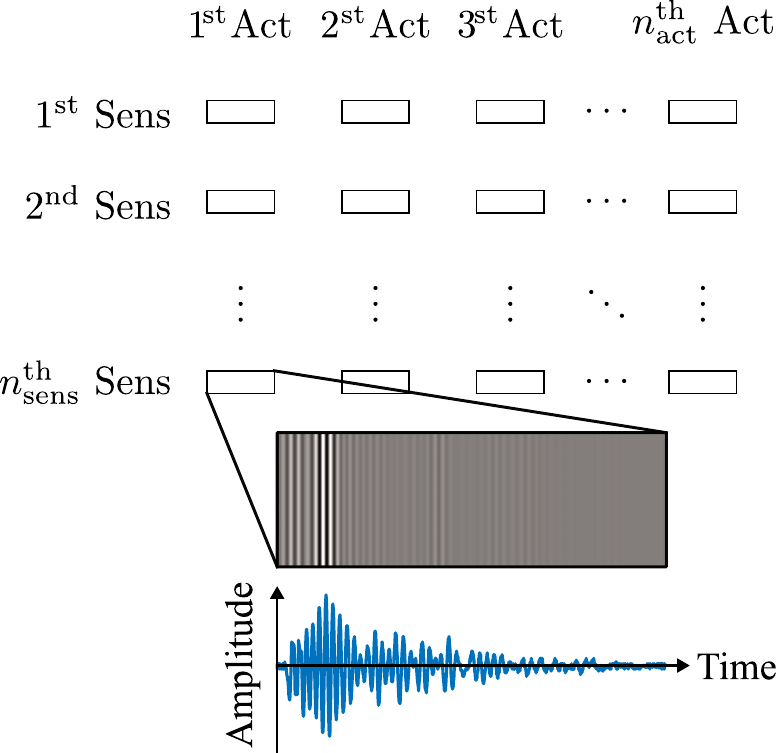}
    \caption{Composition of greyscale image for a pitch-catch acquisition.}
    \label{fig:imageComp}
\end{figure}

The acquired signals were then processed to generate the datasets for the CNNs.
This process is usually referred to as data encoding, and the encoding technique should be able to maintain all the local and global features of the signals while keeping the computational cost low.
The Grayscale (GS) encoding approach \cite{GS} was adopted in this study.
Each time series was normalised according to global maximum and minimum values across the whole dataset, and converted into an image with dimension 1 $\times$ 1,321, where 1,321 identifies the number of data points in the time series.
After conversion, all the signals sensed during the same pitch-catch acquisition were stacked to generate the GSI corresponding to that acquisition, according to the guidelines provided in Ref.~\cite{LOMAZZI2023106003}.
In this context, a pitch-catch acquisition refers to the collection of excitations from all transducers, specifically pertaining to a single damage scenario.
Particularly, the UGW excited by the i-th actuator and sensed by the j-th sensor were placed in the ij position of the GSI, as shown in Figure \ref{fig:imageComp}.
Hence, each GSI had dimensions $n_{sens} \times n_{act}\cdot 1,321$, where $n_{sens}$ is the number of sensors and $n_{act}$ is the number of actuators.
Therefore, the resulting image has a dimension of $7$ pixels $\times 10,568$ pixels.
Each dataset contained 32 images for the rectangular array and 16 images for the circular array.
That is, each dataset included a GS image per damage position.
Then, to improve training performance, data augmentation was employed by adding randomly distributed Gaussian noise to the images \cite{gauss} with a minimum signal-to-noise ratio of 20 $dB$, as suggested in the literature \cite{Ewald,app10238610}.
This allowed enlarging each dataset by 100 times.

\subsection{Damage Localisation}\label{sec:damLoc}
Damage localisation was performed using CNNs for regression.
Three different architectures were employed to handle the varying shapes of the input data, which changed with the application of the TL process.
Tables \ref{tab:CNNfull}-\ref{tab:CNNtl_rect} report the CNN architectures.
It is possible to notice that the output of the CNNs has been fixed to 1, as it represents either the x or y coordinate.
On the contrary, the input dimension is defined as a function of the input image dimension, as it changes from one case study to another due to the dimensionality reduction obtained.
Type-1 CNNs were used to localise damage before the use of MPCA.
Instead, Type-2 and Type-3 CNNs were employed after MPCA to process data from circular and rectangular sensor arrays, respectively.
As can be seen in Tables \ref{tab:CNNfull} and \ref{tab:CNNtl_rect}, the only difference between the feed-forward parts of the CNNs is the presence of a fully connected layer of 30 neurons for Type-1 CNNs. This is due to the higher dimensionality of the output of the convolutional part of the network.
Based on the idea that the way CNNs learn to extract features in the source domain remains valid and transferable to the target domain, only the feed-forward part of the CNNs is fine-tuned in this work.


\begin{table}[hbtp]
\centering
\caption{Architecture of the Type-1 CNNs.}
\adjustbox{width=\columnwidth}{
\begin{tabular}{c|c|c}
\textbf{\#} & \textbf{Layer} & \textbf{Parameters} \\
\hline
1 & Input & imageInputLayer([H, W, CS]) \\
2 & Convolutional 2D & filter size: [1, 6], 4 filters, stride: [1, 3] \\
3 & Batch Normalisation & - \\
4 & ReLU & - \\
5 & Convolutional 2D & filter size: [1, 4], 4 filters, stride: [1, 2] \\
6 & Batch Normalisation & - \\
7 & ReLU & - \\
8 & MaxPooling2D & size: [1, 6], stride: [1, 2] \\
9 & Convolutional 2D & filter size: [1, 2], 32 filters \\
10 & Batch Normalisation & - \\
11 & ReLU & - \\
12 & Convolutional 2D & filter size: [1, 4], 4 filters, stride: [1, 2] \\
13 & Batch Normalisation & - \\
14 & MaxPooling2D & size: [1, 6], stride: [1, 2] \\
15 & Convolutional 2D & filter size: [1, 2], 32 filters \\
16 & Batch Normalisation & - \\
17 & ReLU & - \\
18 & Dropout & - \\
\hline
19 & Fully Connected & 30 neurons \\
20 & Sigmoid & - \\
21 & Fully Connected & 20 neurons \\
22 & Sigmoid & - \\
23 & Fully Connected & 10 neurons \\
24 & Sigmoid & - \\
25 & Fully Connected & 5 neurons \\
26 & Sigmoid & - \\
27 & Fully Connected & 1 neuron \\
28 & Regression & - \\
\end{tabular}
}
\label{tab:CNNfull}
\end{table}

\begin{table}[hbtp]
\centering
\caption{Architecture of the Type-2 CNNs.}
\adjustbox{width=\columnwidth}{
\begin{tabular}{c|c|c}
\textbf{\#} & \textbf{Layer} & \textbf{Parameters} \\
\hline
1 & Input & imageInputLayer([H, W, CS]) \\
2 & Convolutional 2D & filter size: [1, 6], 4 filters, stride: [1, 3] \\
3 & Batch Normalisation & - \\
4 & ReLU & - \\
5 & Convolutional 2D & filter size: [2, 4], 4 filters, stride: [1, 2] \\
6 & Batch Normalisation & - \\
7 & ReLU & - \\
8 & MaxPooling2D & size: [2, 6], stride: [1, 2] \\
9 & Convolutional 2D & filter size: [2, 2], 32 filters \\
10 & Batch Normalisation & - \\
11 & ReLU & - \\
12 & Dropout & - \\
\hline
13 & Fully Connected & 20 neurons \\
14 & Sigmoid & - \\
15 & Fully Connected & 10 neurons \\
16 & Sigmoid & - \\
17 & Fully Connected & 5 neurons \\
18 & Sigmoid & - \\
19 & Fully Connected & 1 neuron \\
20 & Regression & - \\
\end{tabular}
}
\label{tab:CNNtl_circ}
\end{table}

\begin{table}[hbtp]
\centering
\caption{Architecture of the Type-3 CNNs.}
\adjustbox{width=\columnwidth}{
\begin{tabular}{c|c|c}
\textbf{\#} & \textbf{Layer} & \textbf{Parameters} \\
\hline
1 & Input & imageInputLayer([H, W, CS]) \\
2 & Convolutional 2D & filter size: [1, 6], 4 filters, stride: [1, 3] \\
3 & Batch Normalisation & - \\
4 & ReLU & - \\
5 & Dropout Layer & - \\
\hline
6 & Fully Connected & 20 \\
7 & Sigmoid & - \\
8 & Fully Connected & 10 \\
9 & Sigmoid & - \\
10 & Fully Connected & 5 \\
11 & Sigmoid & - \\
12 & Fully Connected & 1 \\
13 & Regression & - \\
\end{tabular}
}
\label{tab:CNNtl_rect}
\end{table}

The input layer processes images with dimensions H x W x CS, where H is the height, W is the width, and CS is the number of channels. In this case, since GS images are used, CS = 1.
The \textit{Batch Normalisation} layer was used to normalise the output of each \textit{Convolutional 2D} layer to obtain a more efficient and stable training. 
\textit{Dropout} layers were used to prevent overfitting.
Two activation functions were used: a \textit{Rectified Linear Unit (ReLU)} for the convolutional part, and a \textit{Sigmoid} for the feedforward part.
\textit{Max Pooling} layers were used to remove redundant information by downsampling the output of \textit{Convolutional 2D} layers.
The feedforward part of the CNNs outputs the prediction through a \textit{Regression} layer with a linear activation function.

Before training, each database was split into a training set, a test set, and a validation set. The training set from the source domain comprised 70\% of the data, while the test and validation sets were constructed with 15\% of the data each. In contrast, fine-tuning was performed by re-training only the feedforward part of the regression CNNs on 90\% of the data, while the test set and validation set each consisted of 5\% of the data.
The different percentages of training data are due to the lower amount of training data for the target domain.

The CNNs were implemented in MATLAB, and training was performed using the Adam optimiser with a batch size of 25, a learning rate of $10^{-3}$, a learning rate drop factor of 0.1, and a learning rate drop period of 15 epochs. Training was stopped at the occurrence of overfitting, or after 50 epochs had elapsed. The network parameters were optimised by minimising the mean square error between the expected and predicted damage positions. Training was conducted on the GPU of an Acer Nitro AN515-57 laptop equipped with an 8-core Intel Core i7-1180 processor, a 6GB NVIDIA GeForce RTX 3060 Laptop GPU, and 16GB of RAM.

The procedure described above was used to train two CNNs for each domain: one to predict the damage position along the $x$ coordinate, and the other to predict the damage position along the $y$ coordinate. The reference system is shown in Figure~\ref{fig:loc}.

\subsection{Domain Adaptation}\label{sec:TL}


This work leverages MPCA on a single database $\mathbf{A}$, which comprises data from both the source domain and the target domain, to perform domain adaptation by identifying shared latent features, e.g., principal components, across the domains.
MPCA is a tensor-to-tensor projection that maximises variations to build a more compact representation of the data, retaining the original tensor structure \cite{MPCA_1,infoMSL}.
As PCA, this technique is generally used to perform feature extraction \cite{MPCA_FE_1,MPCA_FE_2} and dimensionality reduction \cite{MPCA_DR_1,MPCA_DR_2,MPCA_DR_3}.
The main difference between MPCA and PCA is that PCA requires tensor data to be vectorised, as it only operates on vectors. However, the transformation from tensors to vectors breaks the relationships between the multiple channels of the tensor, leading to information loss \cite{MPCA_1,infoMSL}.
For this reason, MPCA was developed to operate directly on tensor objects, preserving the relationships among their channels \cite{MPCA_1}.\\
%

\begin{figure*}[htbp]
    \centering
    \includegraphics[width=0.8\textwidth]{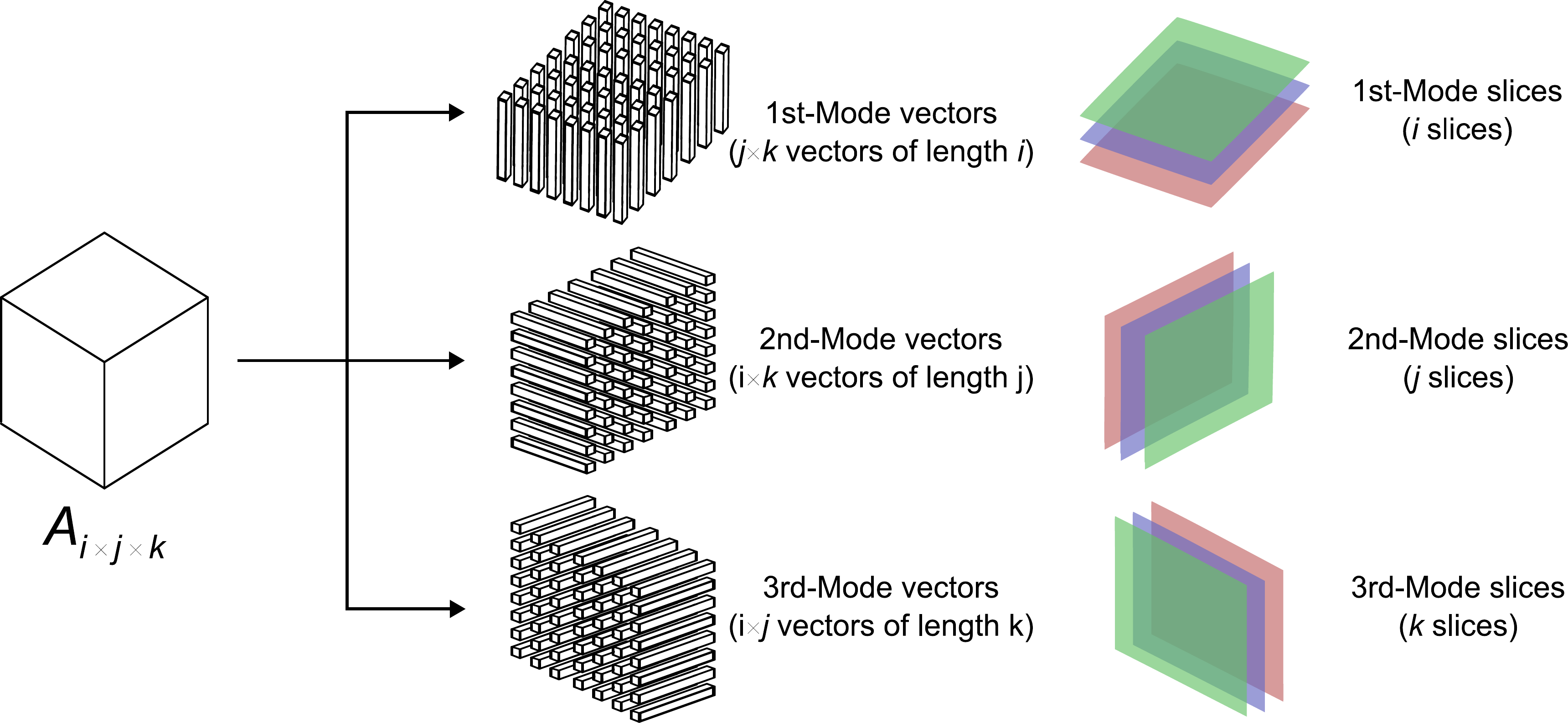}
    \caption{Example of tensor decomposition for a generic 3D case.}
    \label{fig:tensorDecomposition}
\end{figure*}

According to the notation used by Lu et al. \cite{MPCA_1}, a tensor $\mathbf{A} \in \mathbb{R}^{\textit{I}_1} \otimes \mathbb{R}^{\textit{I}_2} \otimes ... \otimes \mathbb{R}^{\textit{I}_n}$ lies in a linear $n$-th order tensor space where $\mathbb{R}^{\textit{I}_1}$, $\mathbb{R}^{\textit{I}_2}$, and $\mathbb{R}^{\textit{I}_N}$ are the linear vectorial spaces, corresponding to $N$ tensor modes \cite{TSA}.
Figure \ref{fig:tensorDecomposition} shows, as an example, the decomposition of a three-dimensional tensor into its three modes, namely 1st-Mode vectors (columns), 2nd-Mode vectors (rows), and 3rd-Mode vectors (tubes).
For high-dimensional data arranged in matrices, it is typically possible to find a lower-dimensional subspace of the original input that captures most of the variation in the data.
Similarly, a tensor subspace can be found for high-dimensional tensors.
To do so, $\textit{P}_n \leq \textit{I}_n$ orthonormal basis vectors are computed for each of the $n$ tensor modes generating a tensor subspace $\mathbb{R}^{\textit{P}_1} \otimes \mathbb{R}^{\textit{P}_2} \otimes ... \otimes \mathbb{R}^{\textit{P}_N}$.
Matrix $\tilde{U}^{(n)}$ can be introduced as the $\textit{I}_n \times \textit{P}_n$ matrix containing the $\textit{P}_n$ $n$-mode basis vectors.
The projection of $\mathbf{A}$ by $\tilde{U}^{(n)}$ can be computed as the inner product between the $n$-mode vector and $\tilde{U}^{(n)^T}$.
The MPCA goal is to find the projection tensor $\tilde{U}^{(n)^T} \in \mathbb{R}^{\textit{I}_n \times \textit{P}_n}$ that maximise the variability held by the reconstructed tensor in all the $N$ modes.

Essentially, MPCA computes the eigenmatrices, i.e., the matrices containing the eigenvectors of each slice for each of the $N$ modes of the tensor decomposition.
Figure \ref{fig:tensorDecomposition} shows the decomposition of a generic 3D tensor $\mathbf{A}$ in slices for each of the $3$ modes.
Thus, there will be three projection tensors $\tilde{U}^{(n)^T}$ in the 3D example represented in Figure \ref{fig:tensorDecomposition}, that is, one per mode.
Each $\tilde{U}^{(n)^T}$ contains a number of eigenmatrices equal to the number of vector decomposition slices.
Referring again to Figure \ref{fig:tensorDecomposition}, the projection tensor $\tilde{U}^{(1)^T}$ will contain $i$ eigenmatrices, one for each 1st-Mode slice.\\
Full projection is obtained by projecting tensor $\mathbf{A}$ through the entire projection tensors $\tilde{U}^{(n)^T}$. This way, the projection of $\mathbf{A}$ holds the whole information and variability of the original tensor $\mathbf{A}$, since $\textit{P}_n = \textit{I}_n$.
However, if a subset $\textit{P}_n < \textit{I}_n$ is chosen, tensor $\mathbf{A}$ is projected into a lower-dimensional subspace, performing dimensionality reduction at the cost of losing some variability.
$\textit{P}_n$ can be selected according to the so-called Q-based method \cite{MPCA_1}, which means that, in similarity with the PCA practice, a percentage of variation to hold is selected.
Then, the least significant eigenvectors are discarded to keep the target variation.
This method is a suboptimal simplification of dimensionality reduction procedures, but leads to similar results compared to more accurate but complex methods, avoiding time-consuming iterative techniques \cite{MPCA_1}.

\begin{figure*}[htbp]
    \centering
    \includegraphics[width=\textwidth]{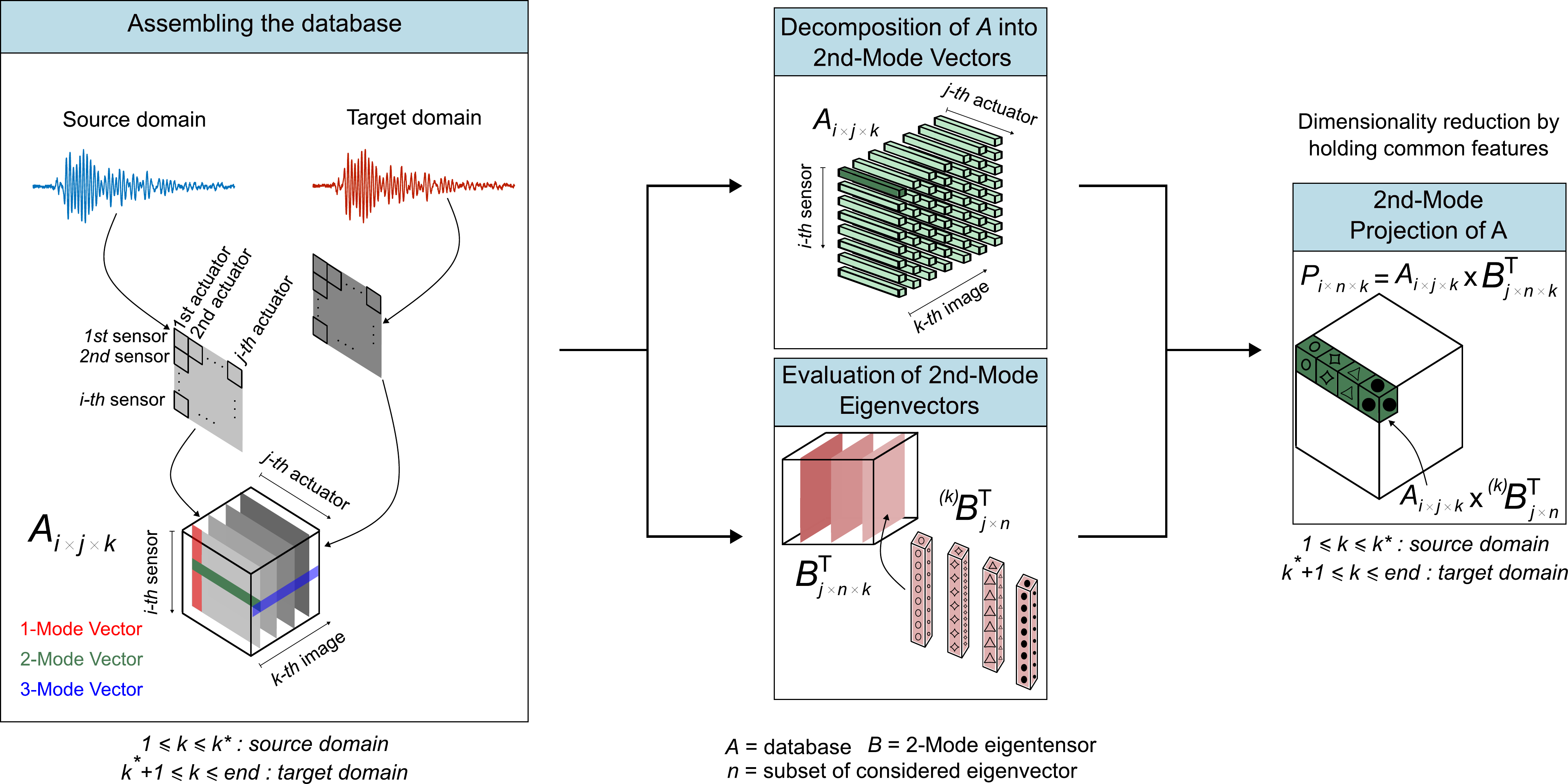}
    \caption{Graphical representation of the MPCA-based domain adaptation method proposed in this work.}
    \label{fig:MPCA_TL}
\end{figure*}

The basic concepts described above were used to develop the MPCA-based domain adaptation approach presented in this work. The proposed method is described in Figure \ref{fig:MPCA_TL}, and is described in the following. With reference to a single case study, a database that encompasses the whole source domain data and half of the target domain data, randomly selected, is generated and represented by a tensor $\mathbf{A}$ .
The tensor database $\mathbf{A}$, schematised in the first block of Figure \ref{fig:MPCA_TL}, consists of:
\begin{itemize}
    \item 2,400 grayscale images ($k$) with 7 rows ($i$), one for each transducer acting as sensor, and 10,568 columns ($j$), in case of circular sensor networks. The first 1,600 images are related to the source domain, while the remaining 800 belong to the target domain.
    \item 4,200 grayscale images ($k$) with 7 rows ($i$), one for each transducer acting as sensor, and 10,568 columns ($j$), in case of rectangular sensor networks. The first 2,800 images are related to the source domain, while the remaining 1,400 belong to the target domain.
\end{itemize}

Tensor $\mathbf{A}$ is decomposed into its 2nd-Mode vectors shown in Figure \ref{fig:MPCA_TL}, and the corresponding 2nd-Mode eigenvectors are computed.
Thus, the projection tensor $\mathbf{B}^T$ related to the 2nd-Mode decomposition is obtained considering the $n$ most relevant eigenvectors.
Eventually, tensor $\mathbf{P}$ is obtained by projecting $\mathbf{A}$ through the projection tensor $\mathbf{B}^T$.
The number of $n$ eigenvectors retained is case study-dependent, and is selected so to hold the desired percentage of variability.
Since MPCA compresses the information along a desired direction, it was applied only to 2nd-Mode vectors (columns) for two reasons: (i) to preserve sensor-related information (rows), and (ii) to avoid data loss by retaining the complete set of available greyscale images (third dimension).
Additionally, 2nd-mode decomposition ensures that every slice contains data from both the source and the target domains.
That is, each slice of the projection tensor $\mathbf{B}^T$ contains eigenvectors related to both domains.
Therefore, this procedure identifies a tensor subspace in which the projected source and target domains are similar to each other, as some of the variability discarded by MPCA is related to differences between the source and target domains.


Statistical metrics were used to quantitatively evaluate the difference in the data distributions before and after MPCA, thereby quantifying the effect of MPCA on data similarity.
Let $H_S$ and $H_T$ be the histograms, e.g., probability distributions, of the source and target data over a common domain, respectively. The common domain ranges from -1 to 1,  since all data are normalised, as described in Section \ref{sec:mat}. The following metrics were considered:

\begin{itemize}
    \item Kullback-Leibler (KL) Divergence $D_{KL}(H_T\parallel H_S)$. This metric was used to evaluate the amount of information lost when $H_S$ is used to approximate $H_T$. The metric is defined according to Equation \ref{eq:KL} \cite{KLD}.
    \begin{equation}
        \centering
        D_{\mathrm{KL}}(H_T \parallel H_S) = \sum_{i=1}^n H_{T,i} \log \frac{H_{T,i}}{H_{S,i}}
        \label{eq:KL}
    \end{equation}
    
    \item Symmetric KL Divergence $D^{sym}_{KL}(H_T\parallel H_S)$. It was used to obtain a more balanced and direction-independent divergence value \cite{KLD}. The metric is defined according to Equation \ref{eq:KLsym} \cite{KLD}.
    \begin{equation}
    \centering
    \begin{split}
    D^{sym}_{\mathrm{KL}}(H_T, H_S) = \frac{1}{2} ( D_{\mathrm{KL}}(H_T \parallel H_S) + \\ + D_{\mathrm{KL}}(H_S \parallel H_T)
    \end{split}
    \label{eq:KLsym}
    \end{equation}

    \item Jensen–Shannon Divergence (JSD). It is a smoothed and symmetric adaptation of the KL divergence defined according to Equation \ref{eq:JSD} \cite{JSD}.
    \begin{equation}
    \centering
    \begin{split}
        JSD(H_T,H_S) = \frac{1}{2} D_{\mathrm{KL}}(H_T \parallel M) + \\ + \frac{1}{2} D_{\mathrm{KL}}(H_S \parallel M) \\ where \quad M = \frac{1}{2}(H_T + H_S)
    \end{split}
    \label{eq:JSD}
    \end{equation}

    \item Chi-Squared Distance $\chi^2(H_T,H_S)$. It is used to measure the discrepancy between two distributions, emphasising relative bin differences \cite{chi}. The metric is defined as shown in Equation \ref{eq:chi},
    \begin{equation}
        \centering
        \chi^2(H_T, H_S) = \sum_{i=1}^n \frac{(H_{T,i} - H_{S,i})^2}{H_{T,i} + H_{S,i}}
        \label{eq:chi}
    \end{equation}

    \item Bhattacharyya Distance $B(H_T,H_S)$. It measures the overlap between two distributions \cite{BD}, with lower values implying higher overlap. Equation \ref{eq:B} shows how this metric is defined.
    \begin{equation}
        \centering
        B(H_T, H_S) = -\ln \left( \sum_{i=1}^n \sqrt{H_{T,i} H_{S,i}} \right)
        \label{eq:B}
    \end{equation}

    \item Earth Mover's Distance (EMD). It is used to evaluate the minimum total work required to transform one distribution into another \cite{emd1,emd2}. Its definition is reported in Equation \ref{eq:emd}. This metric represents the shift between domains, i.e., the extent by which the probability mass of one distribution must be relocated to represent another distribution.
    \begin{equation}
        \centering
        EMD = \sum_{i=1}^{n} \left| \sum_{j=1}^{i} (H_{T,i} - H_{S,i}) \right|
        \label{eq:emd}
    \end{equation}
\end{itemize}

Since these statistical metrics measure the distance between the distributions of the source and target domains, the lower their value, the closer the distributions are, and thus the more similar they are.

\subsection{Procedure}\label{sec:Procedure}
\begin{figure*}[hbtp]
    \centering
    \includegraphics[width=0.8\linewidth]{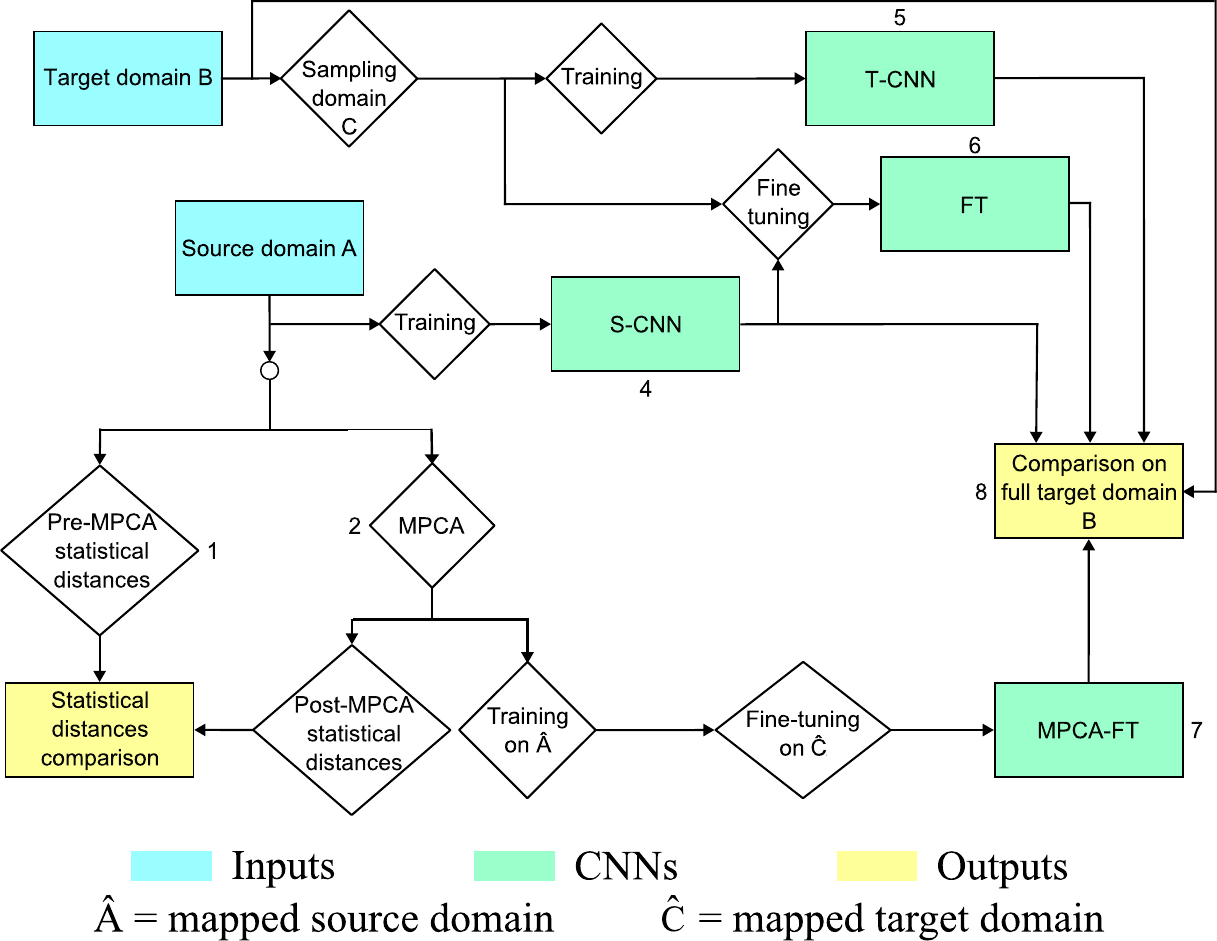}
    \caption{Schematic of the procedure employed in this work. The numbers associated with each block are consistent with the numbering used in the description of the procedure in Section~\ref{sec:Procedure}}
    \label{fig:schematic}
\end{figure*}

The workflow presented in this work is summarised below and schematised in Figure \ref{fig:schematic}.

First, the data from source domain \textit{A} and target domain \textit{B} need to be selected.
All data in the source domain is considered. The target domain has been divided into two randomly selected halves by randomly selecting half of the $k$ grayscale images.
One half is retained and forms a new dataset \textit{C}, while the other is discarded. This is done to simulate data scarcity for the target domain.
Then, the procedure is as follows:
\begin{enumerate}
    \item Computation of statistical metrics to evaluate the similarity of distributions \textit{A} and \textit{C}.
    \item Application of MPCA, retaining an adequate amount of variability. Typically, 99\% is used. Domains \textit{A} and \textit{C} are mapped into the new domains \textit{$\hat{A}$} and \textit{$\hat{C}$}.
    \item Re-evaluation of statistical distance between reduced distributions to quantify the effect of MPCA on data similarity.
    \item Training of two CNNs for regression on database \textit{A}: one to predict the $x$ coordinate of the damage (\textit{S-CNN-x}), and the other for the $y$ coordinate (\textit{S-CNN-y}). The combination of these networks for damage localisation is referred to as \textit{S-CNN}.
    \item Training of two CNNs for regression on database \textit{C}: one to predict the $x$ coordinate of the damage (\textit{T-CNN-x}), and the other for the $y$ coordinate (\textit{T-CNN-y}). \textit{T-CNN} is the name used to identify the combination of these CNNs for damage localisation.
    \item Fine-tuning \textit{S-CNN-x} and \textit{S-CNN-y} on the target domain \textit{C}, obtaining networks \textit{FT-x} and \textit{FT-y}. The term \textit{FT} is used to refer to the combination of these networks to obtain predictions on the damage positions.
    \item Training of two CNNs for regression, one to predict the $x$ coordinate of the damage and the other for the $y$ coordinate, on the mapped source domain \textit{$\hat{A}$}. Fine-tuning of these networks on the mapped target domain \textit{$\hat{C}$} to obtain networks \textit{MPCA-FT-x} and \textit{MPCA-FT-x}, whose combination is named as \textit{MPCA-FT}.
    \item Testing the performance of all networks on the full target domain \textit{B}.
\end{enumerate}
The numbering presented here is consistent with that reported in Figure~\ref{fig:schematic}.

\section{Case Studies}\label{sec:caseStudy}
The proposed framework was evaluated on two tasks: (i) adapt a CNN trained on a material to another material, given the same sensor network layout, and (ii) adapt a CNN trained on a sensor network with a given layout to another layout, without changing the material.
In the following, the two tasks are referred to as \textit{Material Adaptation} and \textit{Sensor Network Adaptation}, respectively.
The former is presented in Section~\ref{sec:changeMat}, while the latter in Section~\ref{sec:changeSN}.


\subsection{Material Adaptation}\label{sec:changeMat}

\subsubsection{Circular Sensor Network}\label{sec:circular}\hfill\\
The following case studies were analysed to demonstrate the capability of the proposed method to handle structures made of different materials, but equipped with the same sensor network:
\begin{itemize}
    \item G16 to K2G4S
    \item K2G4S to G16
    \item G16 to K8
    \item K8 to G16
    \item K8 to K2G4S
    \item K2G4S to K8
\end{itemize}
Each case study is described below in a dedicated sub-Section.

\paragraph{G16 to K2G4S}\label{sec:G16toK2G8Sc}\hfill\\
%
%
Each grayscale image fed into the CNNs has dimensions $7 \times 10,568$. After applying MPCA, the dimensionality is reduced to $7 \times 3,283$ and $7 \times 62$, retaining 99\% and 97\% of the total variability, respectively.
The significant dimensionality reduction achieved when discarding just 1–3\% of the variability indicates that a large number of principal components contributes minimally to the overall variance.

Table \ref{tab:Circ_G16_to_K2G8S_dist} summarizes the statistical analysis results used to assess the similarity between the source and target domain data distributions.
All metrics indicate a reduction in the distance between the distributions after applying MPCA while retaining 99\% of the information. In contrast, reducing the retained variability to 97\% results in an increased distance between the source and target domains, regardless of the metric used. This may be attributed to the excessive dimensionality reduction from 99\% to 97\%, which could lead to excessive discretisation of the distributions.
Thus, the results associated with the 97\% dimensionality reduction are not presented in the following.

\begin{table}[htbp]
    \centering
    \caption{Statistical analysis results used to assess the similarity between the source (G16) and target (K2G4S) domain data distributions.}
    \label{tab:Circ_G16_to_K2G8S_dist}
    \adjustbox{width=\columnwidth}{
    \begin{tabular}{l|ccc}
        \begin{tabular}{c} Statistical \\ Distance\end{tabular} & No MPCA & 99\% MPCA & 97\% MPCA \\
        \hline
        $KL$ & $2.47 \times 10^{-2}$ & $6.30 \times 10^{-3}$ & $4.41 \times 10^{-1}$ \\
        $KL_{sym}$ & $1.33 \times 10^{-2}$ & $6.20 \times 10^{-3}$ & $3.97 \times 10^{-1}$ \\
        $JSD$ & $6.10 \times 10^{-4}$ & $2.94 \times 10^{-4}$ & $1.66 \times 10^{-2}$ \\
        $\chi^2$ & $2.00 \times 10^{-3}$ & $9.03 \times 10^{-4}$ & $5.13 \times 10^{-2}$ \\
        $B$ & $8.02 \times 10^{-4}$ & $4.00 \times 10^{-4}$ & $2.25 \times 10^{-2}$ \\
        $EMD$ & $8.93 \times 10^{-5}$ & $8.42 \times 10^{-6}$ & $4.97 \times 10^{-4}$ \\
    \end{tabular}
    }
\end{table}

\begin{figure}[htbp]
    \centering
    \includegraphics[width=0.8\columnwidth]{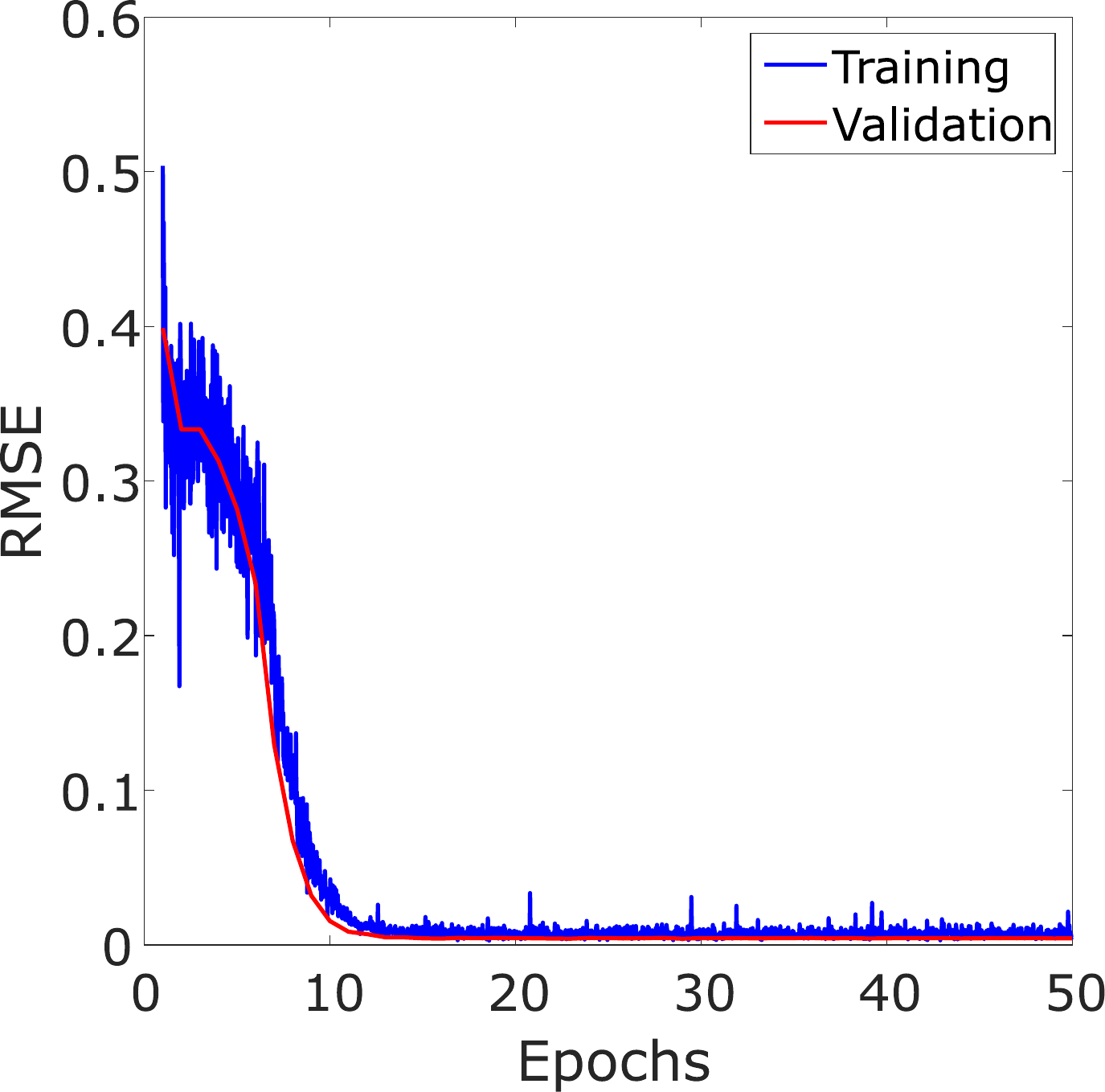}
    \caption{Training performance of \textit{S-CNN-x}.}
    \label{fig:exampleTraining}
\end{figure}

The training performance for the CNN \textit{S-CNN-x} is shown in Figure \ref{fig:exampleTraining}.
All other CNNs exhibit the same global behaviour, and their training performance is omitted for brevity.
Convergence is reached within 50 epochs without overfitting.

\begin{figure}[htbp]
    \centering
    \includegraphics[width=\columnwidth]{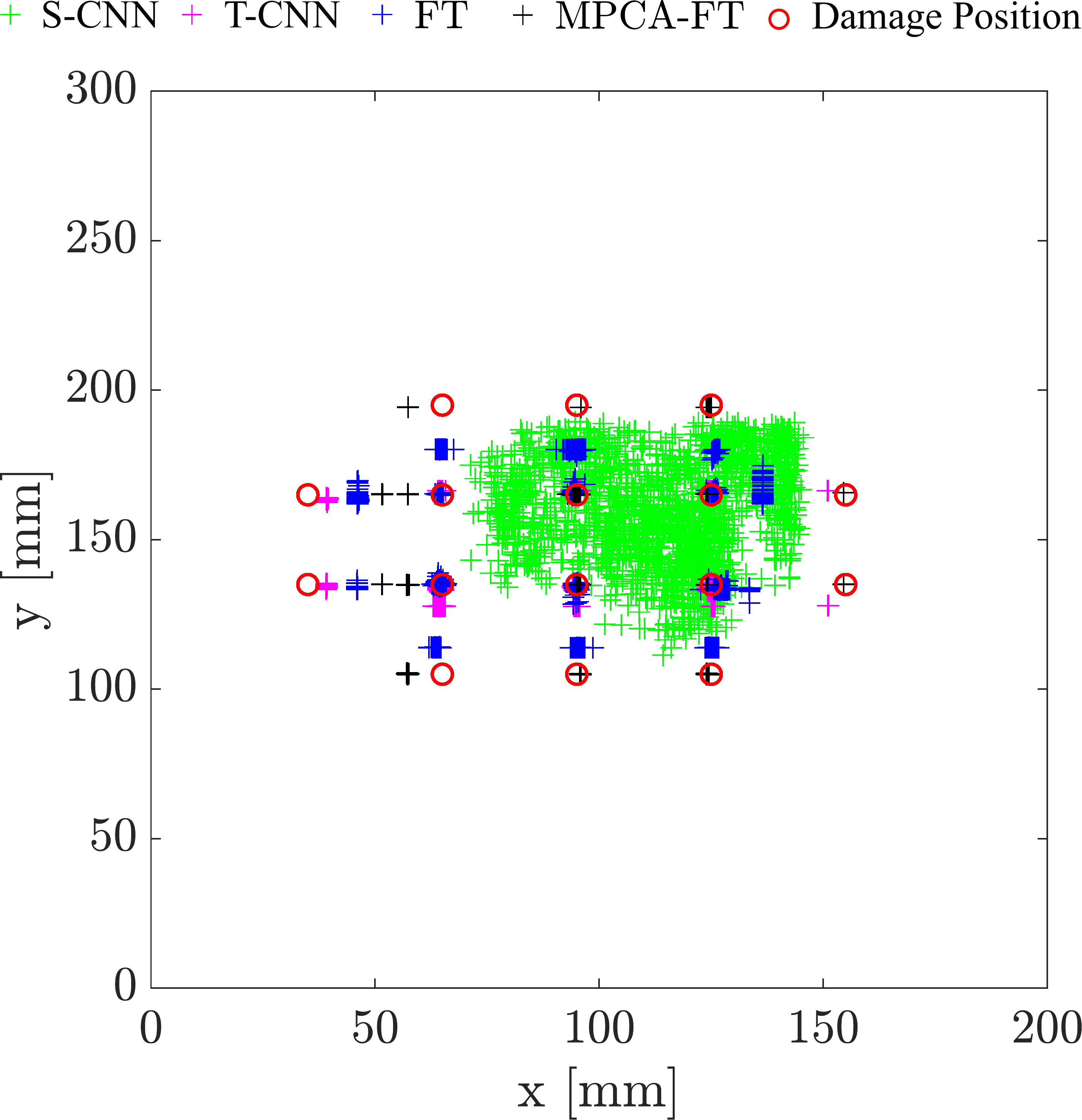}
    \caption{Damage positions predicted by all CNNs.}
    \label{fig:Circ_G16_K2G8S}
\end{figure}

The damage positions predicted by all CNNs are shown in Figure \ref{fig:Circ_G16_K2G8S}. 
In the figure, crosses $\rm{+}$ are used for predictions, and red circles \textcolor{red}{$\rm{\circ}$} for the expected damage positions.
As expected, the CNNs trained on the source domain do not provide accurate predictions on the target domain.
Specifically, the predicted locations \textcolor{green}{$\rm{+}$} are mostly clustered near the centre of the scanning area, deviating from the actual damage positions \textcolor{red}{$\rm{\circ}$}.
Similarly, the CNNs trained on the reduced target domain \textit{C} (\textcolor{magenta}{$\rm{+}$}) cannot accurately localise damage due to data scarcity.
After fine-tuning, localisation accuracy improves slightly, as the predictions \textcolor{blue}{$\rm{+}$} begin to shift toward their corresponding expected positions.
However, the results remain unsatisfactory, as only the central damage locations are accurately captured.
The accuracy further improves after applying MPCA.
In fact, the MPCA-FT networks yield satisfactory predictions \textcolor{black}{$\rm{+}$}, closely matching all the expected damage locations.
A quantitative analysis of the prediction accuracy is provided in Table~\ref{tab:Circ_G16_to_K2G8S_error}, reporting the root mean square localisation error along the $x$ and $y$ axes.
The MPCA-FT networks provide the smallest prediction error, confirming the potentiality of the proposed framework.


\begin{table}[htbp]
    \centering
    \caption{Root mean square localisation error along the $x$ and $y$ axes for each CNN configuration, evaluated on the target (K2G4S) domain data. Lower values indicate better localisation performance.}
    \label{tab:Circ_G16_to_K2G8S_error}
    \begin{tabular}{l|cc}
        & X Error [$\mathrm{mm}$] & Y Error [$\mathrm{mm}$] \\
        \hline
        S-CNN & 33.56 & 33.95 \\
        T-CNN           & 18.55 & 11.79 \\
        FT               & 24.83 & 0.67  \\
        MPCA-FT        & 7.95  & 0.27  \\
    \end{tabular}
\end{table}


\paragraph{K2G4S to G16}\label{sec:K2G8StoG16c}\hfill\\
%
%
The application of MPCA reduces the size of the GSIs to 7$\times$3,272 and 7$\times$62, retaining 99\% and 97\% of the variability, respectively.
Table \ref{tab:Circ_K2G8S_to_G16_dist} shows the statistical metrics before and after MPCA. It is evident that excessive dimensionality reduction, caused by retaining 97\% of the information, is detrimental to the similarity of the distributions between the source and target domains.
In contrast, using 99\% MPCA only removes the differences between the distributions, leading to an increase in similarity and overlap.
However, it can be observed that the $KL$ distance increases from $1.90 \times 10^{-3}$ to $4.77 \times 10^{-3}$ when applying 99\% MPCA.
This behaviour can be justified by the $KL$ divergence's sensitivity to asymmetries and support mismatches, as its symmetric version, $KL_{sym}$, decreases by one order of magnitude, as also done by $JSD$.
Despite the behaviour of $KL$, also in this case study 99\% MPCA improves data similarity, while 97\% MPCA makes it worse.


\begin{table}[htbp]
    \centering
    \caption{Statistical metrics to assess the similarity between source (K2G4S) and target (G16) domain data distributions.}
    \label{tab:Circ_K2G8S_to_G16_dist}
    \adjustbox{width=\columnwidth}{
    \begin{tabular}{l|ccc}
        \begin{tabular}{c} Statistical \\ Distance\end{tabular} & No MPCA & 99\% MPCA & 97\% MPCA \\
        \hline
        $KL$ & $1.90 \times 10^{-3}$ & $4.77 \times 10^{-3}$ & $3.14 \times 10^{-1}$ \\
        $KL_{sym}$ & $1.32 \times 10^{-2}$ & $4.92 \times 10^{-3}$ & $3.48 \times 10^{-1}$ \\
        $JSD$ & $6.07 \times 10^{-4}$ & $2.39 \times 10^{-4}$ & $1.35 \times 10^{-2}$ \\
        $\chi^2$ & $1.90 \times 10^{-3}$ & $7.21 \times 10^{-4}$ & $4.07 \times 10^{-2}$ \\
        $B$ & $7.97 \times 10^{-4}$ & $3.28 \times 10^{-4}$ & $1.88 \times 10^{-2}$ \\
        $EMD$ & $8.89 \times 10^{-5}$ & $4.01 \times 10^{-6}$ & $2.41 \times 10^{-4}$ \\
    \end{tabular}
    }
\end{table}

\begin{figure}[htbp]
    \centering
    \includegraphics[width=\columnwidth]{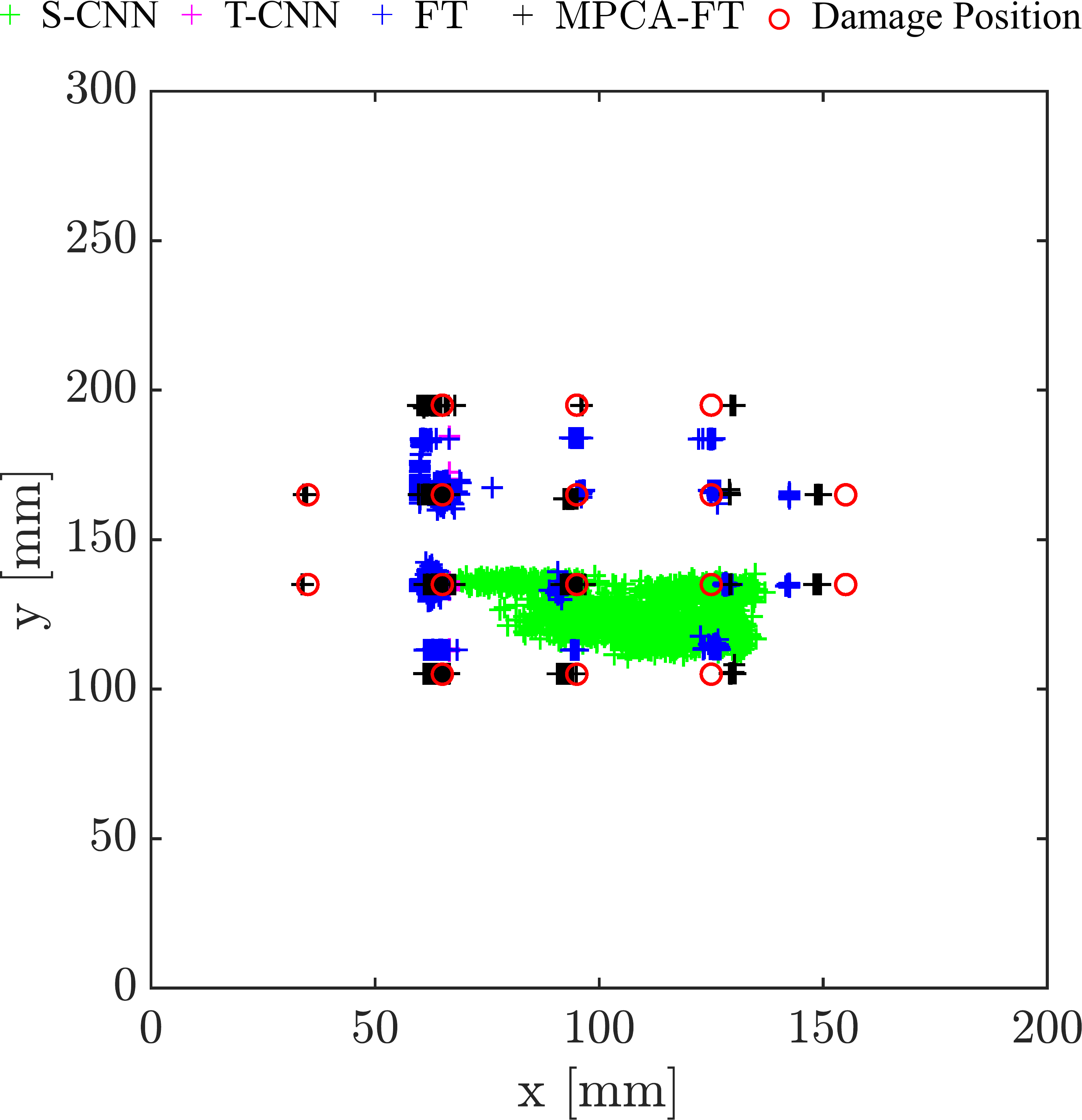}
    \caption{Damage positions predicted by all CNNs.}
    \label{fig:Circ_K2G8S_G16}
\end{figure}

The networks predictions are shown in Figure \ref{fig:Circ_K2G8S_G16}, while Table \ref{tab:Circ_K2G8S_to_G16_error} presents the average prediction error on the $x$ and $y$ axes for the four CNNs.
The predictions of \textit{S-CNN} \textcolor{green}{$\rm{+}$} are all localised to the centre of the plate, highlighting that the networks have not learned any damage-related pattern in the GSIs.
On the contrary, \textit{T-CNN} (\textcolor{magenta}{$\rm{+}$}) learns to predict the $y$ coordinate of damage positions, but not the $x$ coordinate.
This behaviour may be related to the more limited variability of damage positions along the $y$ coordinate.
Fine-tuning is beneficial both with ($\rm{+}$) and without  (\textcolor{blue}{$\rm{+}$}) MPCA application.
Without MPCA, fine-tuning (\textit{FT}) yields a considerable improvement in predicting the actual damage position, with errors of 10.10 $mm$ and 6.26 $mm$ in the horizontal and vertical locations, respectively.
Combining MPCA and fine-tuning (\textit{MPCA-FT}) further improves prediction accuracy, as black crosses ($\rm{+}$) go closer to the expected damage positions, as also evidenced by the smaller prediction errors.\\


\begin{table}[htbp]
    \centering
    \caption{Root mean square localisation error along the $x$ and $y$ axes for each CNN configuration, evaluated on the target (G16) domain data. Lower values indicate better localisation performance.}
    \label{tab:Circ_K2G8S_to_G16_error}
    \begin{tabular}{l|cc}
        & X Error [$\mathrm{mm}$] & Y Error [$\mathrm{mm}$] \\
        \hline
        S-CNN & 55.08 & 36.23 \\
        T-CNN & 46.44 & 5.69  \\
        FT & 10.10 & 6.26  \\
        MPCA-FT & 3.46  & 0.37  \\
    \end{tabular}
\end{table}


\paragraph{G16 to K8}\label{sec:G16toK8c}\hfill\\
%
%
Applying 99\% MPCA reduces the dimensionality of the GSIs to $7\times3,808$.
Notably, this scenario retains a higher number of dimensions compared to the previous two cases.
This is likely attributable to the composition of the panels: $K2G4S$ is a hybrid of $G16$ and $K8$, and its behaviour is expected to resemble both materials more closely than $G16$ and $K8$ resemble each other.
Table \ref{tab:Circ_G16_to_K8_dist} presents the statistical distances between the data distributions before and after applying MPCA.
In all cases, MPCA reduces the statistical distance between the $G16$ and $K8$ distributions, indicating that it not only preserves but enhances their similarity.
This effect arises because the discarded 1\% of variance contains components related to features that are not shared between the distributions.


\begin{table}[hbtp]
    \centering
    \caption{Statistical metrics to assess the similarity between source (G16) and target (K8) domain data distributions.}
    \label{tab:Circ_G16_to_K8_dist}
    \begin{tabular}{l|cc}
        \begin{tabular}{c} Statistical \\ Distance\end{tabular} & No MPCA & 99\% MPCA \\
        \hline
        $KL$ & $1.07 \times 10^{-2}$ & $3.40 \times 10^{-3}$\\
        $KL_{sym}$ & $6.11 \times 10^{-3}$ & $3.53 \times 10^{-3}$\\
        $JSD$ & $4.57 \times 10^{-4}$ & $1.85 \times 10^{-4}$\\
        $\chi^2$ & $1.61 \times 10^{-3}$ & $5.68 \times 10^{-4}$\\
        $B$ & $5.37 \times 10^{-4}$ & $2.48 \times 10^{-4}$\\
        $EMD$ & $9.28 \times 10^{-5}$ & $4.29 \times 10^{-5}$\\
    \end{tabular}
\end{table}

\begin{figure}[htbp]
    \centering
    \includegraphics[width=\columnwidth]{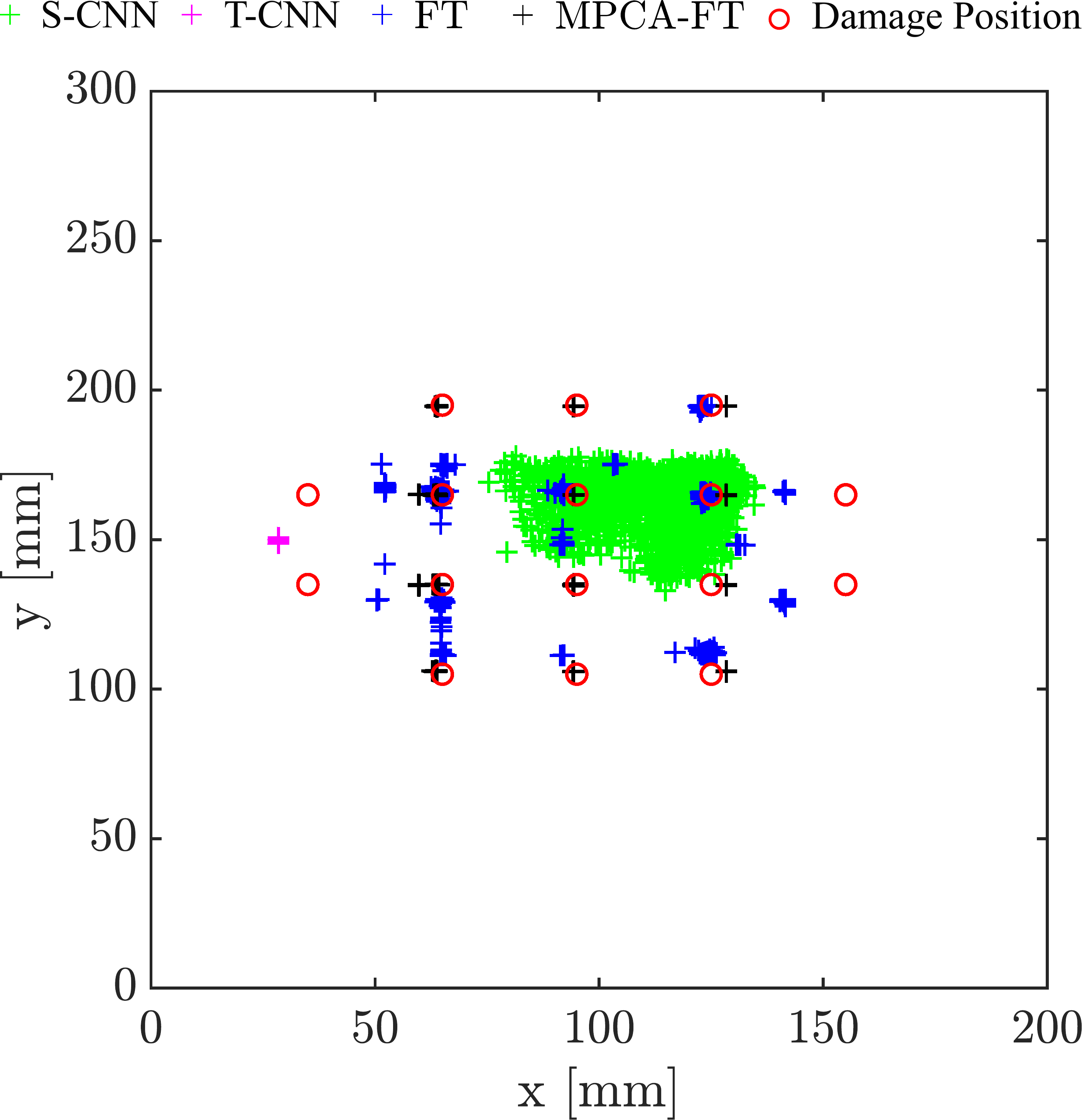}
    \caption{Damage positions predicted by all CNNs.}
    \label{fig:Circ_G16_K8}
\end{figure}

Figure \ref{fig:Circ_G16_K8} shows the predictions of the four networks.
The damage positions estimated by the \textit{S-CNN} (\textcolor{green}{$\rm{+}$}) are randomly scattered around the centre of the plate.
In contrast, the \textit{T-CNN} predictions (\textcolor{magenta}{$\rm{+}$}) are clustered toward the left side of the plate.
The \textit{FT} and \textit{MPCA-FT} approaches yield more reliable results, as their predicted damage locations (\textcolor{blue}{$\rm{+}$} and $\rm{+}$, respectively) are closer to the actual damage positions (\textcolor{red}{$\rm{\circ}$}).
As shown in Table \ref{tab:Circ_G16_to_K8}, fine-tuning improves the regression-based localisation performance of the CNNs, regardless of whether MPCA is applied.
However, MPCA specifically enhances the accuracy of the $y$ coordinate, reducing its prediction error from 9.46 $mm$ to 0.44 $mm$.
Conversely, the error in the $x$ coordinate increases from 8.14 $mm$ to 12.99 $mm$ with MPCA.
As illustrated in Figure \ref{fig:Circ_G16_K8}, this increase is due to mispredictions occurring only at the outer horizontal damage locations in the network trained with MPCA.
For central damage positions, however, localisation accuracy slightly improves.


\begin{table}[htbp]
    \centering
    \caption{Root mean square localisation error along the $x$ and $y$ axes for each CNN configuration, evaluated on the target (K8) domain data. Lower values indicate better localisation performance.}
    \label{tab:Circ_G16_to_K8}
    \begin{tabular}{l|cc}
        & X Error [$\mathrm{mm}$] & Y Error [$\mathrm{mm}$] \\
        \hline
        S-CNN       & 44.26 & 33.95 \\
        T-CNN       & 76.02 & 29.31 \\
        FT           & 8.14  & 9.46  \\
        MPCA-FT    & 12.99 & 0.44  \\
    \end{tabular}
\end{table}

\paragraph{K8 to G16}\label{sec:K8toG16c}\hfill\\
This case study reverses the analysis presented in Section \ref{sec:G16toK8c}, with MPCA producing the same dimensionality reduction and resulting in a final GSI size of $7\times3,808$.
As shown in Table \ref{tab:Circ_K8_to_G16_dist}, the symmetric Kullback–Leibler divergence ($KL_{sym}$) decreases from $6.09 \times 10^{-3}$ to $3.83 \times 10^{-3}$.
Both the Jensen–Shannon divergence ($JSD$) and the Bhattacharyya distance ($B$) are more than halved by MPCA, while the $\chi^2$ and Earth Mover’s Distance ($EMD$) drop by an order of magnitude, indicating improved alignment and similarity between distributions.
However, the standard Kullback–Leibler divergence ($KL$) more than doubles after MPCA.
While this may seem contradictory, it can be attributed to the nature of the $KL$ metric, which is asymmetric and highly sensitive to mismatches in the distribution tails: small discrepancies in low-probability regions can lead to large increases in its value.
Therefore, symmetric and bounded measures such as $JSD$, $KL_{sym}$, and $EMD$ provide a more stable and reliable assessment of distributional similarity.
This highlights that $KL$ alone may not offer a comprehensive evaluation when comparing dimensionality-reduced distributions.


\begin{table}[hbtp]
    \centering
    \caption{Statistical metrics to assess the similarity between source ($K8$) and target ($G16$) domain data distributions.}
    \label{tab:Circ_K8_to_G16_dist}
    \begin{tabular}{l|cc}
        \begin{tabular}{c} Statistical \\ Distance\end{tabular} & No MPCA & 99\% MPCA \\
        \hline
        $KL$ & $1.54 \times 10^{-3}$ & $3.95 \times 10^{-3}$\\
        $KL_{sym}$ & $6.09 \times 10^{-3}$ & $3.83 \times 10^{-3}$\\
        $JSD$ & $4.56 \times 10^{-4}$ & $1.81 \times 10^{-4}$\\
        $\chi^2$ & $1.61 \times 10^{-3}$ & $5.45 \times 10^{-4}$\\
        $B$ & $5.36 \times 10^{-4}$ & $2.49 \times 10^{-4}$\\
        $EMD$ & $9.27 \times 10^{-5}$ & $4.41 \times 10^{-6}$\\
    \end{tabular}
\end{table}

\begin{figure}[htbp]
    \centering
    \includegraphics[width=\columnwidth]{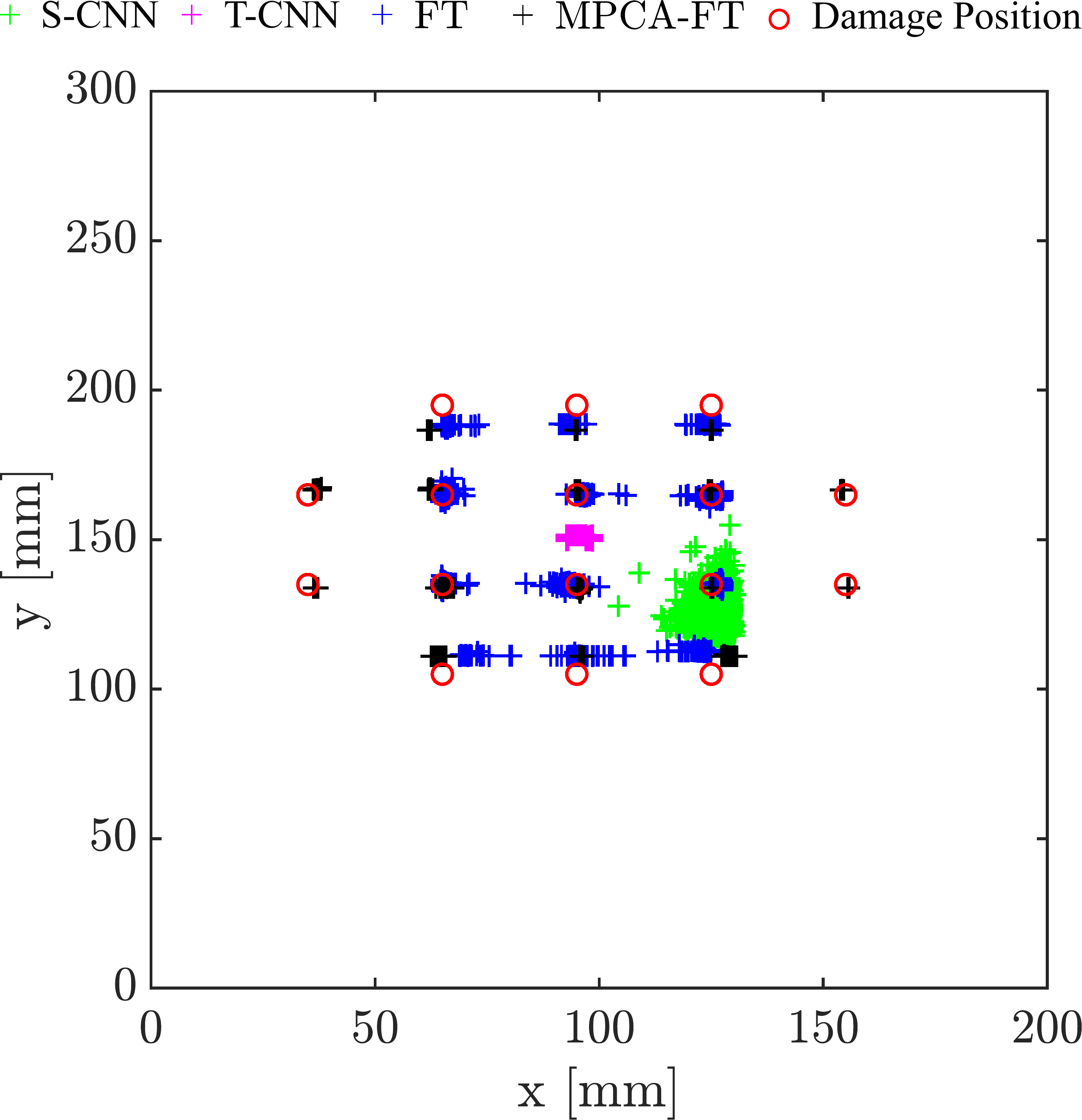}
    \caption{Damage positions predicted by all CNNs.}
    \label{fig:Circ_K8_G16}
\end{figure}

Figure \ref{fig:Circ_K8_G16} illustrates the predicted damage locations produced by the different CNN models, while Table \ref{tab:Circ_K8_to_G16} reports their corresponding prediction errors.
In the figure, the predictions from \textit{S-CNN} (\textcolor{green}{$\rm{+}$}) are clustered in the right corner of the monitored area.
In contrast, those from \textit{T-CNN} (\textcolor{magenta}{$\rm{+}$}) are concentrated near the centre.
Fine-tuning enhances localisation accuracy primarily for central damage positions, indicating that it does not fully resolve the distribution mismatch between the source and target domains.
In contrast, the combination of fine-tuning and MPCA leads to a substantial improvement in prediction accuracy: the black crosses ($\rm{+}$) from the \textit{MPCA-FT} model are consistently closest to the true damage locations.
This observation is supported by the results in Table \ref{tab:Circ_K8_to_G16}, where the RMSE of the \textit{MPCA-FT} predictions is 3.82 $mm$ for the $x$ coordinate and 1.21 $mm$ for the $y$ coordinate. \\




\begin{table}[htbp]
    \centering
    \caption{Root mean square localisation error along the $x$ and $y$ axes for each CNN configuration, evaluated on the target (G16) domain data. Lower values indicate better localisation performance.}
    \label{tab:Circ_K8_to_G16}
    \begin{tabular}{l|cc}
        & X Error [$\mathrm{mm}$] & Y Error [$\mathrm{mm}$] \\
        \hline
        S-CNN       & 37.04 & 38.03 \\
        T-CNN       & 36.58 & 25.56 \\
        FT           & 17.78 & 5.59  \\
        MPCA-FT    & 3.82  & 1.21  \\
    \end{tabular}
\end{table}

\paragraph{K8 to K2G4S}\label{sec:K8toK2G8Sc}\hfill\\
MPCA reduces the dimensionality of the database to $7 \times 3,296$ while retaining 99\% of the total variance.
Unlike previous cases, MPCA produces a distinct behaviour here, as shown in Table \ref{tab:Circ_K8_to_K2G8S_dist}.
The Kullback–Leibler divergence ($KL$) increases from $4.37 \times 10^{-4}$ to $3.00 \times 10^{-3}$, and the Jensen–Shannon divergence ($JSD$) also rises from $9.24 \times 10^{-5}$ to $1.19 \times 10^{-4}$.
Similarly, the symmetric $KL$ divergence ($KL_{sym}$) increases from $3.83 \times 10^{-4}$ to $2.41 \times 10^{-3}$, and the Bhattacharyya distance ($B$) nearly doubles, suggesting that MPCA introduces distortions affecting the overlap between distributions.
In contrast, the $\chi^2$ distance remains nearly unchanged, and the Earth Mover’s Distance ($EMD$) decreases by an order of magnitude, indicating improved alignment in terms of mass distribution.
This mixed behaviour may be explained by the fact that MPCA introduces small mismatches in low-probability regions, which increase divergence-based metrics such as $KL$, $KL_{sym}$, $JSD$, and $B$.
The consistency of this effect across multiple metrics suggests these mismatches are systematic.
Nonetheless, the absolute values of all metrics remain very small, implying that the original distributions were already highly similar.
Thus, despite the relative increases in some metrics, the numerical impact is negligible.
Moreover, the reduction in $EMD$ indicates a better geometric alignment between distributions, even if minor perturbations appear in their probability densities.
At first glance, these results may seem to contradict those of Sections \ref{sec:G16toK8c} and \ref{sec:K8toG16c}, where MPCA improved similarity between materials $G16$ and $K8$.
However, in those cases, MPCA was applied between two plates made of distinct materials, whereas $K2G4S$ is a hybrid of $G16$ and $K8$, possibly introducing more complex or broader feature distributions.
This difference is reflected in the number of retained features: both the $G16 \rightarrow K8$ and $K8 \rightarrow G16$ cases retain 3,808 features, while dimensionality drops to 3,283 for $G16 \rightarrow K2G4S$ and 3,272 for the reverse.
In the $K8 \rightarrow K2G4S$ case, the dimension is slightly higher (3,296), but overall, the reductions involving $K2G4S$ suggest that fewer features are needed to preserve 99\% of the variance.
This implies that $K2G4S$ shares more features with both $G16$ and $K8$, enabling MPCA to discard more variance without compromising representativeness.
The relatively small differences in retained dimensionality — 3,283 for $G16 \rightarrow 2G4S$, 3,272 for $K2G4S \rightarrow G16$, and 3,296 for $K8 \rightarrow K2G4S$ — are likely due to variations in the number of shared features depending on whether a material is used as the source or target domain, which in turn determines whether all or only a subset of the data is considered.


\begin{table}[hbtp]
    \centering
    \caption{Statistical metrics to assess the similarity between source ($K8$) and target ($K2G4S$) domain data distributions.}
    \label{tab:Circ_K8_to_K2G8S_dist}
    \begin{tabular}{l|cc}
        \begin{tabular}{c} Statistical \\ Distance\end{tabular} & No MPCA & 99\% MPCA \\
        \hline
        $KL$ & $4.37 \times 10^{-4}$ & $3.00 \times 10^{-3}$\\
        $KL_{sym}$ & $3.83 \times 10^{-4}$ & $2.41 \times 10^{-3}$\\
        $JSD$ & $9.24 \times 10^{-5}$ & $1.19 \times 10^{-4}$\\
        $\chi^2$ & $3.57 \times 10^{-4}$ & $3.65 \times 10^{-4}$\\
        $B$ & $9.41 \times 10^{-5}$ & $1.62 \times 10^{-4}$\\
        $EMD$ & $1.83 \times 10^{-5}$ & $1.35 \times 10^{-6}$\\
    \end{tabular}
\end{table}

\begin{figure}[htbp]
    \centering
    \includegraphics[width=\columnwidth]{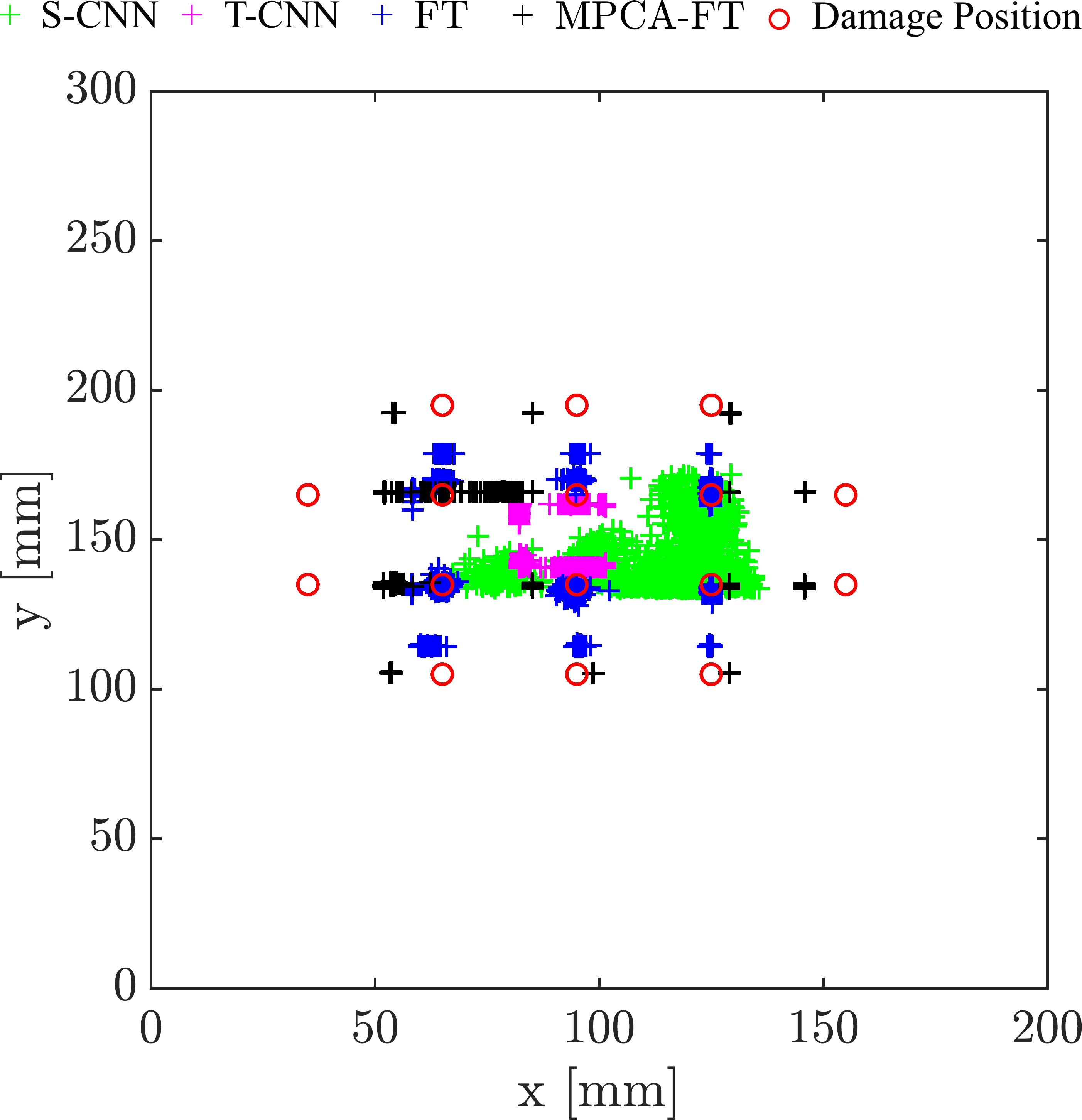}
    \caption{Damage positions predicted by all CNNs.}
    \label{fig:Circ_K8_K2G8S}
\end{figure}

Figure \ref{fig:Circ_K8_K2G8S} presents the estimated damage locations produced by the different networks, while Table \ref{tab:Circ_K8_to_K2G8S} reports their corresponding prediction errors along the $x$ and $y$ axes.
Both \textit{S-CNN} (\textcolor{green}{$\rm{+}$}) and \textit{T-CNN} (\textcolor{magenta}{$\rm{+}$}) fail to accurately localise the damage, with their predictions scattered across the central region of the plate.
This poor performance is reflected in the high prediction errors shown in Table \ref{tab:Circ_K8_to_K2G8S}.
Fine-tuning improves localisation accuracy, as the \textit{FT} model yields more reliable predictions (\textcolor{blue}{$\rm{+}$}) that are closer to the true damage locations.
However, \textit{FT} still struggles to detect damage located at the plate’s outer regions, as also indicated by the error metrics in Table \ref{tab:Circ_K8_to_K2G8S}.
By contrast, the \textit{MPCA-FT} model achieves the most accurate localisation, with predicted positions ($\rm{+}$) closely matching the actual damage locations.
The only exceptions are the two vertical damages on the left side of the plate, which are not correctly identified.
Nevertheless, the model successfully localises the two vertical damages on the right side.
This improvement is supported by the error values reported in Table \ref{tab:Circ_K8_to_K2G8S}, where \textit{MPCA-FT} achieves prediction errors of 10.12 $mm$ and 1.31 $mm$ along the $x$ and $y$ axes, respectively.



\begin{table}[htbp]
    \centering
    \caption{Root mean square localisation error along the $x$ and $y$ axes for each CNN configuration, evaluated on the target (K2G4S) domain data. Lower values indicate better localisation performance.}
    \label{tab:Circ_K8_to_K2G8S}
    \begin{tabular}{l|cc}
        & X Error [$\mathrm{mm}$] & Y Error [$\mathrm{mm}$] \\
        \hline
        S-CNN       & 42.46 & 34.39 \\
        T-CNN       & 29.23 & 21.49 \\
        FT           & 13.43 & 8.39  \\
        MPCA-FT    & 10.12 & 1.31  \\
    \end{tabular}
\end{table}

\paragraph{K2G4S to K8}\label{sec:K2G8StoK8c}\hfill\\
%
%
Retaining 99\% of the variance results in a dimensionality reduction to $7 \times 3,292$.
As shown in Table \ref{tab:Circ_K2G8S_to_K8_dist}, the behavior of the metrics mirrors that observed in the previous section.
Divergence-based measures—including $KL$, $KL_{sym}$, $JSD$, and $B$—increase, with both the asymmetric and symmetric $KL$ values showing a rise of nearly an order of magnitude.
Although these increases are proportionally large, their absolute values remain very small, indicating that the distributions remain highly similar after MPCA.
The $\chi^2$ distance increases only slightly, from $3.57 \times 10^{-4}$ to $4.04 \times 10^{-4}$.
In contrast, the Earth Mover’s Distance ($EMD$) improves by nearly an order of magnitude, further supporting that MPCA enhances alignment in terms of mass distribution.


\begin{table}[hbtp]
    \centering
    \caption{Statistical metrics to assess the similarity between source ($K2G4S$) and target ($K8$) domain data distributions.}
    \label{tab:Circ_K2G8S_to_K8_dist}
    \begin{tabular}{l|cc}
        \begin{tabular}{c} Statistical \\ Distance\end{tabular} & No MPCA & 99\% MPCA \\
        \hline
        $KL$ & $3.27 \times 10^{-4}$ & $2.93 \times 10^{-3}$\\
        $KL_{sym}$ & $3.82 \times 10^{-4}$ & $2.96 \times 10^{-3}$\\
        $JSD$ & $9.22 \times 10^{-5}$ & $1.35 \times 10^{-4}$\\
        $\chi^2$ & $3.57 \times 10^{-4}$ & $4.04 \times 10^{-4}$\\
        $B$ & $9.39 \times 10^{-5}$ & $1.87 \times 10^{-4}$\\
        $EMD$ & $1.82 \times 10^{-5}$ & $2.86 \times 10^{-6}$\\
    \end{tabular}
\end{table}

\begin{figure}[htbp]
    \centering
    \includegraphics[width=\columnwidth]{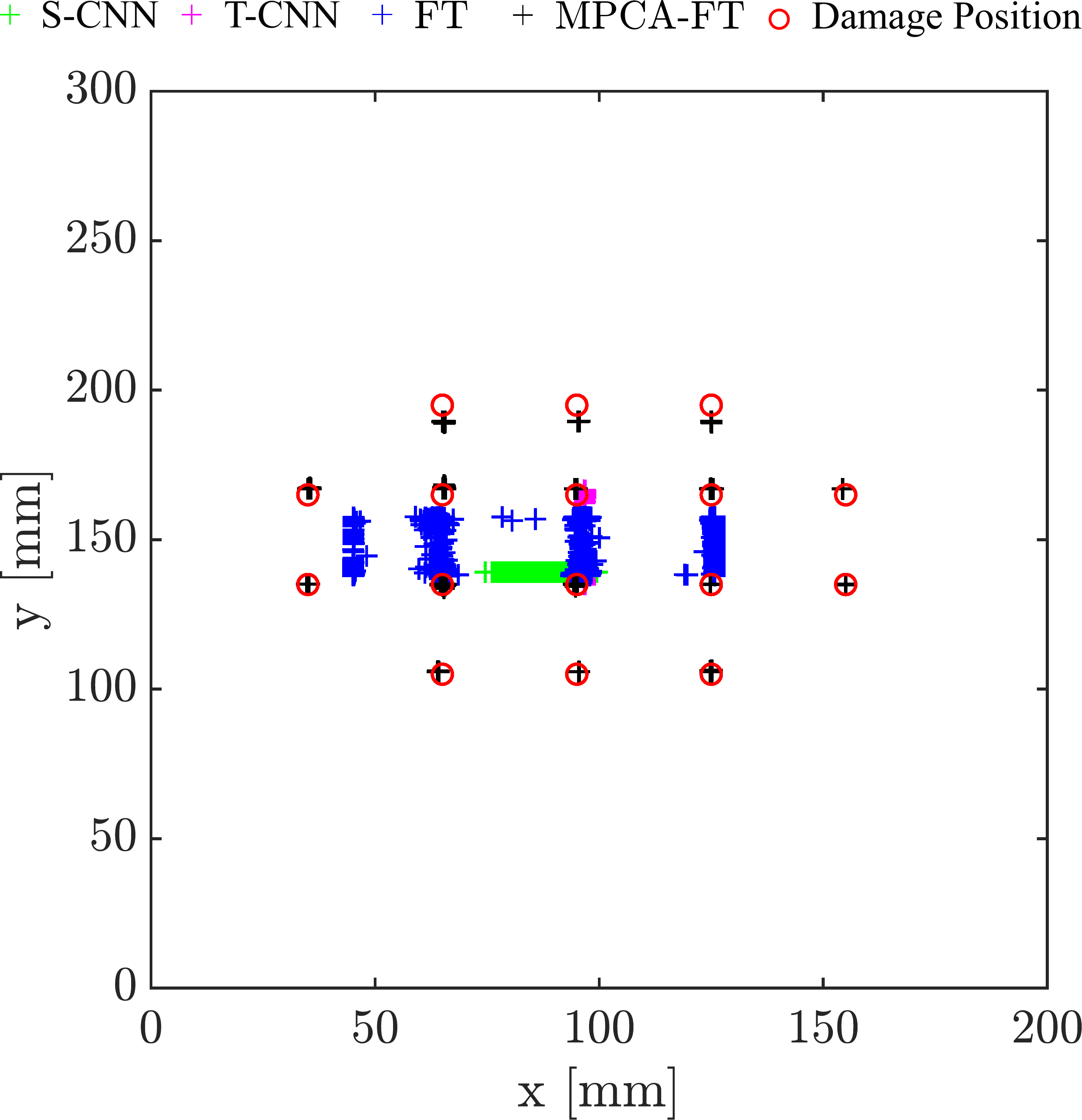}
    \caption{Damage positions predicted by all CNNs.}
    \label{fig:Circ_K2G8S_K8}
\end{figure}

The analysis of the results is based on Figure \ref{fig:Circ_K2G8S_K8} and Table \ref{tab:Circ_K2G8S_to_K8}.
Both \textit{S-CNN} and \textit{T-CNN} fail to provide accurate damage localisation, as evidenced by the scattered predictions (\textcolor{green}{$\rm{+}$} and \textcolor{magenta}{$\rm{+}$}, respectively) in Figure \ref{fig:Circ_K2G8S_K8}.
This is corroborated by the high prediction errors reported in Table \ref{tab:Circ_K2G8S_to_K8}.
The \textit{FT} model demonstrates a slight improvement, with predicted positions (\textcolor{blue}{$\rm{+}$}) showing marginally better alignment, but still failing to achieve reliable accuracy.
This is reflected in the persistently high errors for both the $x$ and $y$ coordinates shown in Table \ref{tab:Circ_K2G8S_to_K8}.
In contrast, the \textit{MPCA-FT} model delivers highly accurate localisation results, with predicted positions ($\rm{+}$) nearly overlapping the true damage locations (\textcolor{red}{$\rm{\circ}$}) in Figure \ref{fig:Circ_K2G8S_K8}.
Accordingly, the lowest prediction errors are observed for \textit{MPCA-FT} in Table \ref{tab:Circ_K2G8S_to_K8}, with values of 0.41 $mm$ and 2.67 $mm$ for the $x$ and $y$ axes, respectively.

\begin{table}[htbp]
    \centering
    \caption{Root mean square localisation error along the $x$ and $y$ axes for each CNN configuration, evaluated on the target (K8) domain data. Lower values indicate better localisation performance.}
    \label{tab:Circ_K2G8S_to_K8}
    \begin{tabular}{l|cc}
        & X Error [$\mathrm{mm}$] & Y Error [$\mathrm{mm}$] \\
        \hline
        S-CNN       & 35.97 & 31.91 \\
        T-CNN       & 36.73 & 17.98 \\
        FT           & 11.15 & 23.18 \\
        MPCA-FT    & 0.41  & 2.67  \\
    \end{tabular}
\end{table}

\subsubsection{Rectangular Sensor Network}\label{sec:rectangular}\hfill\\
This section presents the case studies where TL is applied across different materials, while maintaining the same rectangular sensor array configuration.
Section \ref{sec:G16toK2G8Sr} explores the adaptation of regression CNNs from the $G16$ dataset to $K2G4S$, while Section \ref{sec:K2G8StoG16r} investigates the reverse direction, transferring from $K2G4S$ to $G16$.

\paragraph{G16 to K2G4S}\label{sec:G16toK2G8Sr}\hfill\\
%
%
The effect of applying 99\% MPCA to the data distributions is analysed in Table \ref{tab:Rect_G16_to_K2G8S_dist}.
In contrast to the results obtained for the circular sensor array (Section \ref{sec:circular}), the table reveals a significant degradation in distributional similarity after MPCA: all divergence metrics increase by approximately two orders of magnitude.
This behaviour can be explained by the extent of dimensionality reduction performed—while the original data has a size of $7 \times 10,568$, it is compressed to $7 \times 39$ after applying 99\% MPCA.
Such a drastic reduction discards a substantial number of original features, potentially including important structural information, despite retaining 99\% of the variance.
From a physics-based perspective, the key difference between this case and the corresponding circular array scenario (Section \ref{sec:G16toK2G8Sc}) lies in the sensor configuration.
The rectangular sensor network enables monitoring of a larger plate area but also results in longer travel paths for the ultrasonic guided waves (UGWs) to reach the sensing PZTs.
This has two important consequences.
First, a larger monitored area implies more damage locations—32 instead of 16—on which MPCA must operate, making the identification of shared features across all cases more challenging.
Second, longer propagation paths lead to increased signal attenuation, further complicating feature extraction.
Together, these factors reduce the effectiveness of MPCA in identifying consistent features across multiple damage cases under high attenuation conditions.
To mitigate this issue, the variance threshold retained by MPCA must be increased.
Raising it to 99.9\% moderately alleviates the compression severity, as shown in the third column of Table \ref{tab:Rect_G16_to_K2G8S_dist}.
With this higher threshold, the data dimension is reduced to $7 \times 119$, which is less aggressive.
This adjustment results in smaller changes in the $KL$ divergence, $JSD$, and $EMD$ relative to the original data, suggesting improved preservation of distributional characteristics.
However, the values of $KL_{sym}$, $\chi^2$, and $B$ still exhibit notable increases, confirming that MPCA continues to discard many low-variance directions that may carry important but subtle structural features.


\begin{table}[hbtp]
    \centering
    \caption{Statistical metrics to assess the similarity between source ($G16$) and target ($K2G4S$) domain data distributions.}
    \label{tab:Rect_G16_to_K2G8S_dist}
    \adjustbox{width=\columnwidth}{
    \begin{tabular}{l|ccc}
        \begin{tabular}{c} Statistical \\ Distance\end{tabular} & No MPCA & 99\% MPCA & 99.9\% MPCA \\
        \hline
        $KL$ & $1.28 \times 10^{-3}$ & $3.27 \times 10^{-1}$ & $9.31 \times 10^{-2}$\\
        $KL_{sym}$ & $8.98 \times 10^{-4}$ & $2.67 \times 10^{-1}$ & $1.00 \times 10^{-1}$\\
        $JSD$ & $1.46 \times 10^{-4}$ & $1.14 \times 10^{-2}$ & $5.10 \times 10^{-3}$\\
        $\chi^2$ & $5.55 \times 10^{-4}$ & $4.43 \times 10^{-2}$ & $1.58 \times 10^{-2}$\\
        $B$ & $1.54 \times 10^{-5}$ & $1.81 \times 10^{-2}$ & $6.68 \times 10^{-3}$\\
        $EMD$ & $5.14 \times 10^{-5}$ & $4.51 \times 10^{-4}$ & $1.38 \times 10^{-4}$\\
    \end{tabular}
    }
\end{table}

\begin{figure}[htbp]
    \centering
    \includegraphics[width=\columnwidth]{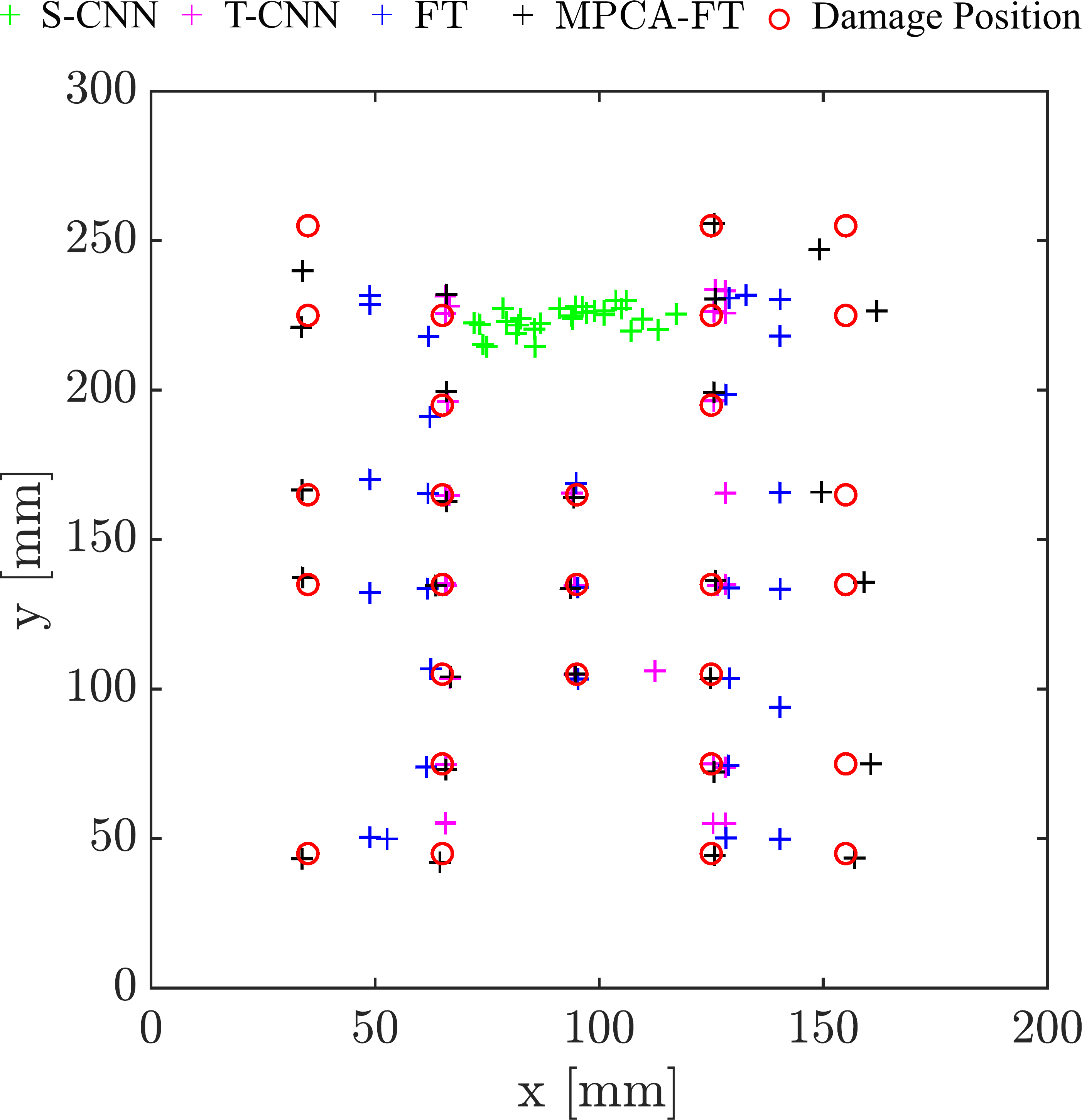}
    \caption{Damage positions predicted by all CNNs.}
    \label{fig:Rect_G16_K2G8S}
\end{figure}

Despite the degradation in distributional similarity metrics, the regression performance of the CNNs—illustrated in Figure \ref{fig:Rect_G16_K2G8S} and detailed in Table \ref{tab:Rect_G16_K2G8S}—demonstrates that 99.9\% MPCA still enables accurate damage localisation.
The black crosses ($\rm{+}$), representing predictions from the \textit{MPCA-FT} model, are consistently closest to the actual damage locations (\textcolor{red}{$\rm{\circ}$}).
In contrast, fine-tuning without MPCA does not achieve comparable performance: the \textit{FT} predictions (\textcolor{blue}{$\rm{\circ}$}) are notably distant from the true damage positions, particularly for damage located on the left and right edges of the plate.
This observation is corroborated by Table \ref{tab:Rect_G16_K2G8S}, which shows significantly lower prediction errors along both the $x$ and $y$ axes for \textit{MPCA-FT} compared to \textit{FT}.
The baseline models, \textit{S-CNN} and \textit{T-CNN}, also perform poorly.
Their predicted damage positions (\textcolor{green}{$\rm{+}$}) are scattered across the upper-central region of the plate, lacking any consistent spatial correlation with the actual damage.
Although \textit{T-CNN} (\textcolor{magenta}{$\rm{+}$}) performs slightly better than \textit{S-CNN}, its predictions remain confined to the plate's central area and are still less accurate than those of the fine-tuned model, as confirmed by the corresponding error values in Table \ref{tab:Rect_G16_K2G8S}.
These findings are particularly noteworthy: although statistical divergence metrics indicate a deterioration in distributional alignment after MPCA, the combination of MPCA with fine-tuning yields better localisation performance than fine-tuning alone.
This suggests that conventional distributional metrics may not fully capture the utility of MPCA in the context of domain adaptation.
Consequently, there may be a need to define complementary or task-specific metrics to evaluate MPCA’s effectiveness in DA frameworks, particularly when it is used alongside parameter-based transfer learning strategies, such as the partial fine-tuning approach adopted in this study.
The improved localisation accuracy achieved through this combination confirms the potential of MPCA to extract transferable features that are beneficial for downstream prediction tasks, even when distributional similarity metrics suggest otherwise.

\begin{table}[htbp]
    \centering
    \caption{Root mean square localisation error along the $x$ and $y$ axes for each CNN configuration, evaluated on the target (K2G4S) domain data. Lower values indicate better localisation performance.}
    \label{tab:Rect_G16_K2G8S}
    \begin{tabular}{l|cc}
        & X Error [$\mathrm{mm}$] & Y Error [$\mathrm{mm}$] \\
        \hline
        S-CNN       & 45.29 & 102.31 \\
        T-CNN       & 18.14 & 8.28 \\
        FT           & 9.63 & 10.58  \\
        MPCA-FT    & 2.60 & 4.12  \\
    \end{tabular}
\end{table}

\paragraph{K2G4S to G16}\label{sec:K2G8StoG16r}\hfill\\
%
%
MPCA was applied using two variance retention thresholds: 90\%, which reduced the image dimensionality to $7 \times 39$, and 99.9\%, which increased it to $7 \times 119$.
Table \ref{tab:Rect_K2G8S_to_G16_dist} reports the statistical distances between the source and target domains before and after MPCA.
Consistent with the findings in Section \ref{sec:G16toK2G8Sr}, a substantial increase in divergence metrics is observed following MPCA, with the most pronounced increases occurring under stronger compression (i.e., 90\% variance retention).
Comparing this case to the previous one, it is evident that significant distributional distortion arises from the use of 99\% MPCA in both scenarios.
However, when $K2G4S$ is used as the source domain—meaning the full $K2G4S$ dataset is retained while only half of the $G16$ data is used—the impact of MPCA-induced compression becomes more pronounced.
This suggests that $K2G4S$ exhibits a broader and more variable data distribution, a hypothesis already supported in Section \ref{sec:K8toK2G8Sc}, which addressed the distinctive behaviours of the three composite plates under the circular sensor network.
The hybrid nature of $K2G4S$, compared to the single-material compositions of $G16$ and $K8$, likely contributes to its more complex distributional characteristics.
This complexity may hinder the MPCA's ability to project $K2G4S$ and $G16$ data into a common low-dimensional subspace without significant information loss, thus explaining the heightened sensitivity to dimensionality reduction in this direction.


\begin{table}[hbtp]
    \centering
    \caption{Statistical metrics to assess the similarity between source ($K2G4S$) and target ($G16$) domain data distributions.}
    \label{tab:Rect_K2G8S_to_G16_dist}
    \adjustbox{width=\columnwidth}{
    \begin{tabular}{l|ccc}
        \begin{tabular}{c} Statistical \\ Distance\end{tabular} & No MPCA &  99\% MPCA & 99.9\% MPCA \\
        \hline
        $KL$ & $5.17 \times 10^{-4}$ & $2.58 \times 10^{-1}$ & $6.87 \times 10^{-2}$\\
        $KL_{sym}$ & $8.95 \times 10^{-4}$ & $2.94 \times 10^{-1}$ & $5.39 \times 10^{-2}$\\
        $JSD$ & $1.45 \times 10^{-4}$ & $1.47 \times 10^{-2}$ & $3.73 \times 10^{-3}$\\
        $\chi^2$ & $5.53 \times 10^{-4}$ & $4.69 \times 10^{-2}$ & $1.23 \times 10^{-2}$\\
        $B$ & $1.54 \times 10^{-4}$ & $1.91 \times 10^{-2}$ & $4.57 \times 10^{-3}$\\
        $EMD$ & $5.13 \times 10^{-5}$ & $5.38 \times 10^{-4}$ & $1.89 \times 10^{-4}$\\
    \end{tabular}
    }
\end{table}

\begin{figure}[htbp]
    \centering
    \includegraphics[width=\columnwidth]{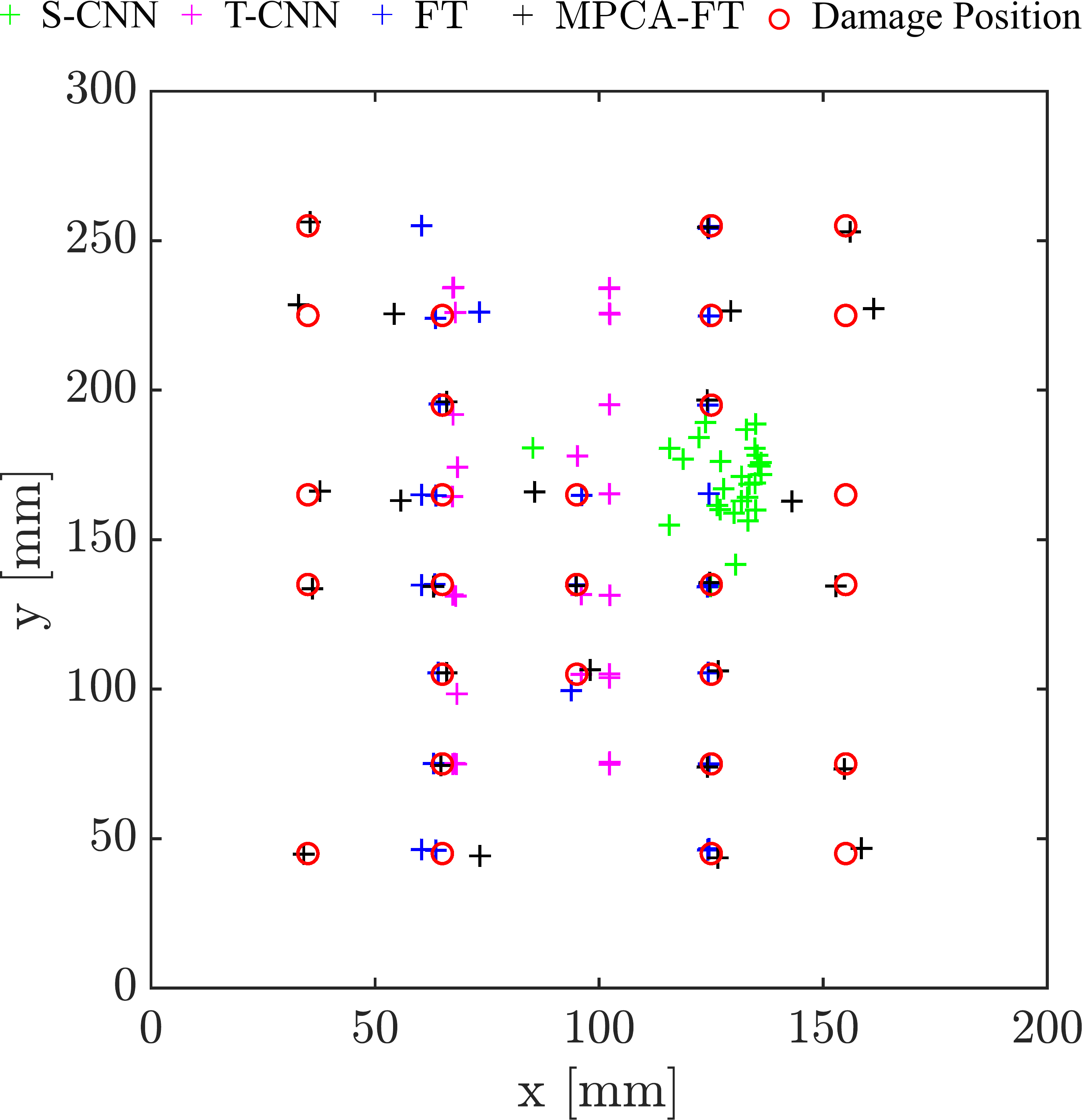}
    \caption{Damage positions predicted by all CNNs.}
    \label{fig:Rect_K2G8S_G16}
\end{figure}

Turning to the performance analysis, Figure \ref{fig:Rect_K2G8S_G16} displays the predicted damage locations across the full target domain, while Table \ref{tab:Rect_K2G8S_to_G16} reports the prediction errors of each regression CNN along the $x$ and $y$ axes.
As in the previous case study, the distortion introduced by dimensionality reduction does not translate into degraded performance.
On the contrary, the combination of 99.9\% MPCA and fine-tuning results in \textit{MPCA-FT} achieving the most accurate and reliable damage localisation.
In Figure \ref{fig:Rect_K2G8S_G16}, the \textit{MPCA-FT} predictions (black crosses, $\rm{+}$) are consistently the closest to the actual damage locations (\textcolor{red}{$\rm{\circ}$}).
By contrast, the \textit{FT} model—fine-tuned without MPCA—fails to reliably identify damage, particularly at the plate’s edges, as shown by the \textcolor{blue}{$\rm{+}$} predictions in the figure.
The \textit{S-CNN} predictions (\textcolor{green}{$\rm{+}$}) are clustered in the upper-central region of the plate and fail to localise any damage positions effectively.
Meanwhile, the \textit{T-CNN} outputs (\textcolor{magenta}{$\rm{+}$}) demonstrate only marginally better performance than \textit{S-CNN}, but still fall short of the accuracy achieved by both fine-tuned models.
These differences in localisation performance are clearly reflected in the prediction errors reported in Table \ref{tab:Rect_K2G8S_to_G16}, where the lowest errors are associated with the \textit{MPCA-FT} model.
This confirms that combining MPCA with fine-tuning is the most effective strategy for enhancing regression-based damage localisation in this domain adaptation scenario.

\begin{table}[htbp]
    \centering
    \caption{Root mean square localisation error along the $x$ and $y$ axes for each CNN configuration, evaluated on the target (G16) domain data. Lower values indicate better localisation performance.}
    \label{tab:Rect_K2G8S_to_G16}
    \begin{tabular}{l|cc}
        & X Error [$\mathrm{mm}$] & Y Error [$\mathrm{mm}$] \\
        \hline
        S-CNN       & 51.36 & 73.50 \\
        T-CNN       & 30.22 & 14.91 \\
        FT           & 18.56 & 1.23  \\
        MPCA-FT    & 4.69 & 1.45  \\
    \end{tabular}
\end{table}

\subsection{Sensor Network Adaptation}\label{sec:changeSN}

\subsubsection{G16}\label{sec:G16}\hfill\\
This section focuses on the knowledge transfer from the circular to the rectangular sensor network, presented in Section \ref{sec:G16Circ2Rect}, holding $G16$ as the shared material.
Section \ref{sec:G16Rect2Circ} describes the inverse process, i.e., the TL from the rectangular sensor array to the circular one.

\paragraph{Circular to Rectangular}\label{sec:G16Circ2Rect}\hfill\\
%
%

The original uncompressed data has a dimensionality of $7 \times 10,568$, which is reduced to $7 \times 1,234$ after applying 99\% MPCA.
To assess the impact of MPCA on data distributions, Table \ref{tab:G16_Circ_to_Rect_dist} provides the relevant statistical metrics.
The table shows that both symmetric and asymmetric divergence measures increase slightly following MPCA.
This trend is also observed for the Jensen–Shannon divergence ($JSD$), the $\chi^2$ distance, and the Bhattacharyya distance ($B$), indicating minor distortions in the probability distributions.
However, the Earth Mover’s Distance ($EMD$) decreases by a factor of two after MPCA, suggesting that the geometric alignment between the distributions actually improves.
Overall, the changes in statistical distances are small, implying that the topology of the sensor network—circular vs. rectangular—has a relatively limited effect on the structure of the data distributions in this case.


\begin{table}[hbtp]
\centering
    \caption{Statistical metrics to assess the similarity between source (circular sensor network) and target (rectangular sensor network) domain data distributions.}
    \label{tab:G16_Circ_to_Rect_dist}
    \begin{tabular}{l|cc}
        \begin{tabular}{c} Statistical \\ Distance\end{tabular} & No MPCA & 99\% MPCA \\
        \hline
        $KL$ & $8.35 \times 10^{-3}$ & $8.67 \times 10^{-3}$\\
        $KL_{sym}$ & $4.60 \times 10^{-3}$ & $7.17 \times 10^{-3}$\\
        $JSD$ & $2.58 \times 10^{-4}$ & $4.32 \times 10^{-4}$\\
        $\chi^2$ & $8.64 \times 10^{-4}$ & $1.40 \times 10^{-3}$\\
        $B$ & $3.25 \times 10^{-4}$ & $5.49 \times 10^{-4}$\\
        $EMD$ & $5.76 \times 10^{-5}$ & $2.39 \times 10^{-5}$\\
    \end{tabular}
\end{table}

\begin{figure}[htbp]
    \centering
    \includegraphics[width=\columnwidth]{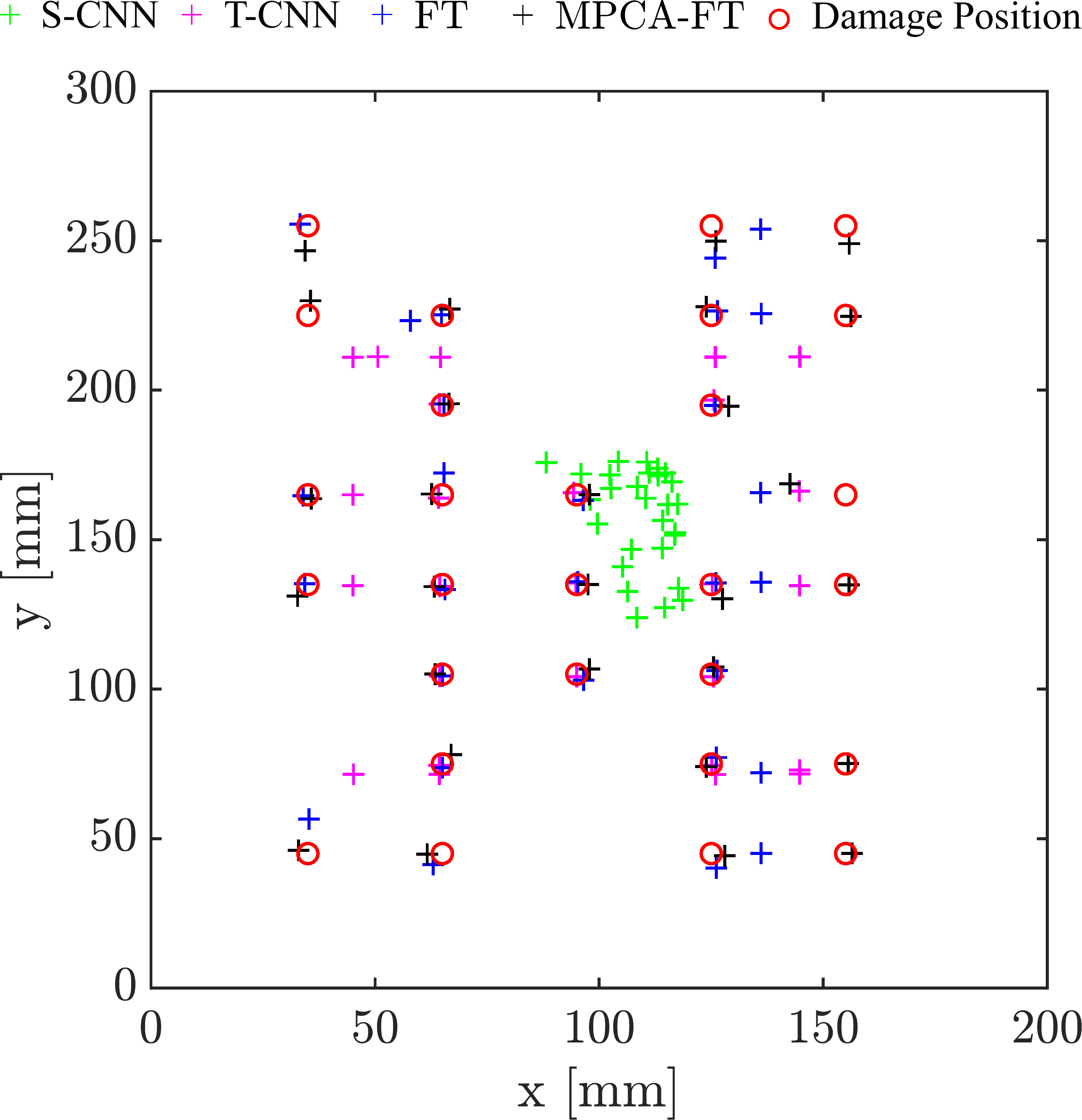}
    \caption{Damage positions predicted by all CNNs.}
    \label{fig:G16_Circ_Rect}
\end{figure}

To compare the performance of the different regression CNNs, Figure \ref{fig:G16_Circ_Rect} illustrates the predicted damage locations on the target domain.
In the figure, \textcolor{green}{$\rm{+}$} denotes the predictions of \textit{S-CNN}, while \textcolor{magenta}{$\rm{+}$} corresponds to \textit{T-CNN}.
Predictions from the \textit{FT} model are shown as \textcolor{blue}{$\rm{+}$}, whereas black crosses ($\rm{+}$) represent the outputs of \textit{MPCA-FT}.
The actual damage locations are indicated by red circles (\textcolor{red}{$\rm{\circ}$}).
The \textit{S-CNN} fails to provide accurate localisation, with predictions scattered far from the true damage sites.
\textit{T-CNN} demonstrates slightly improved performance and performs comparably to the \textit{FT} model.
Both models are able to localise damage located near the centre of the plate, but fail to correctly identify external damage positions.
In contrast, \textit{MPCA-FT} delivers the most reliable predictions, with its outputs closely aligned with the actual damage locations.
These qualitative observations from Figure \ref{fig:G16_Circ_Rect} are confirmed by the quantitative localisation errors reported in Table \ref{tab:G16_circ_to_rect}.
The table clearly shows that the combination of MPCA and fine-tuning results in the lowest prediction errors among all models.
Conversely, using fine-tuning alone — or training the regression CNNs solely on the source or target domain — leads to significantly less accurate localisation outcomes.

\begin{table}[htbp]
    \centering
    \caption{Root mean square localisation error along the $x$ and $y$ axes for each CNN configuration, evaluated on the target (rectangular sensor array) domain data. Lower values indicate better localisation performance.}
    \label{tab:G16_circ_to_rect}
    \begin{tabular}{l|cc}
        & X Error [$\mathrm{mm}$] & Y Error [$\mathrm{mm}$] \\
        \hline
        S-CNN       & 45.10 & 71.14 \\
        T-CNN       & 6.79 & 18.31 \\
        FT           & 9.80 & 3.66  \\
        MPCA-FT    & 3.04 & 2.96  \\
    \end{tabular}
\end{table}

\paragraph{Rectangular to Circular}\label{sec:G16Rect2Circ}\hfill\\
%
%

The original data has a dimensionality of $7 \times 10,568$, which is reduced to $7 \times 110$ after applying 99
The more aggressive compression observed in this case study — compared to that discussed in Section \ref{sec:G16Circ2Rect} — may be attributed to the different objective of the analysis, which likely influenced the feature retention pattern during dimensionality reduction.
This compression is accompanied by a significant degradation in the metrics used to assess data similarity before and after MPCA, as shown in Table \ref{tab:G16_Rect_to_Circ_dist}.
The table indicates increases of one to two orders of magnitude across all statistical distance measures, suggesting that MPCA introduces a notable distortion in the data distribution when projecting the data into a lower-dimensional space.
This highlights a potential trade-off between dimensionality reduction and distributional integrity, especially in scenarios where structural or domain-specific variability plays a critical role.


\begin{table}[hbtp]
    \centering
    \caption{Statistical metrics to assess the similarity between source (rectangular sensor network) and target (circular sensor network) domain data distributions.}
    \label{tab:G16_Rect_to_Circ_dist}
    \begin{tabular}{l|cc}
        \begin{tabular}{c} Statistical \\ Distance\end{tabular} & No MPCA & 99\% MPCA \\
        \hline
        $KL$ & $8.48 \times 10^{-4}$ & $1.92 \times 10^{-1}$ \\
        $KL_{sym}$ & $4.61 \times 10^{-3}$ & $1.76 \times 10^{-1}$ \\
        $JSD$ & $2.60 \times 10^{-4}$ & $7.78 \times 10^{-3}$ \\
        $\chi^2$ & $8.69 \times 10^{-4}$ & $2.38 \times 10^{-2}$ \\
        $B$ & $3.27 \times 10^{-5}$ & $1.05 \times 10^{-2}$ \\
        $EMD$ & $5.78 \times 10^{-5}$ & $2.45 \times 10^{-4}$\\
    \end{tabular}
\end{table}

\begin{figure}[htbp]
    \centering
    \includegraphics[width=\columnwidth]{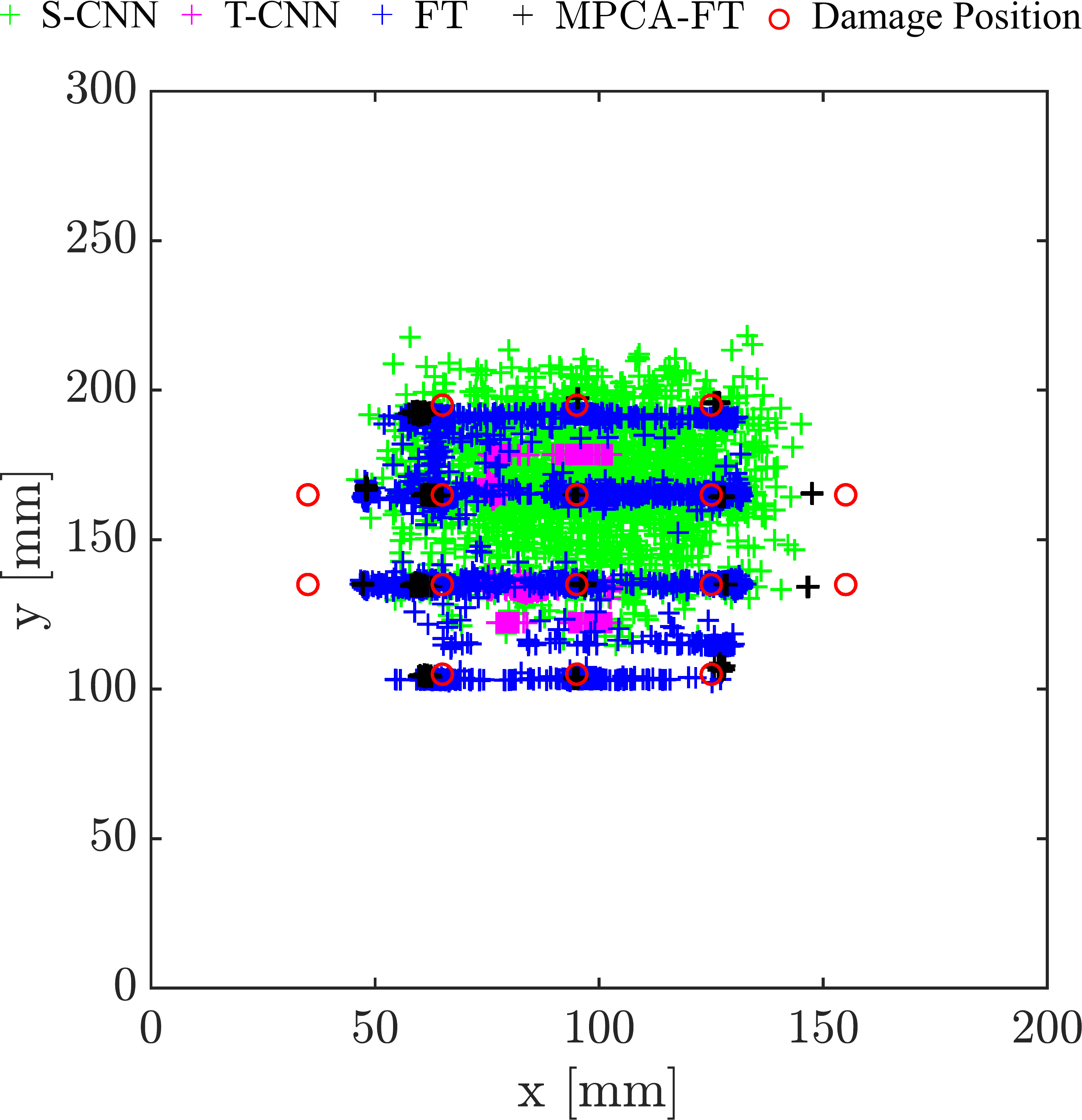}
    \caption{Damage positions predicted by all CNNs.}
    \label{fig:G16_Rect_Circ}
\end{figure}

Figure \ref{fig:G16_Rect_Circ} demonstrates that the best damage localisation performance is achieved by the \textit{MPCA-FT} model, as indicated by the close alignment between its predicted positions ($\rm{+}$) and the actual damage locations (\textcolor{red}{$\rm{\circ}$}).
In contrast, all other models fail to provide reliable predictions, with their outputs appearing randomly distributed across the plate.
This qualitative observation is quantitatively confirmed by the prediction errors reported in Table \ref{tab:G16_rect_to_circ}, where \textit{MPCA-FT} yields the lowest errors for both the $x$ and $y$ directions.
These results align well with the discussion presented in Section \ref{sec:rectangular}, which addressed transfer learning across different materials using the rectangular sensor network.
In both cases, the findings highlight the limitations of relying solely on statistical distance metrics to assess the effectiveness of MPCA in DA scenarios.
However, a key distinction lies in the degree of dimensionality reduction required.
Unlike the case discussed in Section \ref{sec:rectangular}, where increasing the retained variance was necessary to preserve model performance, here the reduction from $7 \times 10,568$ to $7 \times 110$ proved sufficient for successful training of the regression CNNs.
For comparison, in Section \ref{sec:rectangular}, the dimensionality was reduced to just $7 \times 39$, indicating that the current configuration allows a better balance between compression and feature preservation.

\vspace{3 cm}

\begin{table}[htbp]
    \centering
    \caption{Root mean square localisation error along the $x$ and $y$ axes for each CNN configuration, evaluated on the target (circular sensor array) domain data. Lower values indicate better localisation performance.}
    \label{tab:G16_rect_to_circ}
    \begin{tabular}{l|cc}
        & X Error [$\mathrm{mm}$] & Y Error [$\mathrm{mm}$] \\
        \hline
        S-CNN       & 42.82 & 40.01 \\
        T-CNN       & 27.14 & 10.38 \\
        FT           & 20.10 & 6.68  \\
        MPCA-FT    & 5.86 & 1.31  \\
    \end{tabular}
\end{table}

\subsubsection{K2G4S}\label{sec:K2G8S}\hfill\\
This section analyses the transfer learning (TL) from one sensor array to another while fixing the material, i.e., $K2G4S$.
In Section \ref{sec:K2G8SCirc2Rect}, the knowledge acquired on the circular sensor array is transferred to the rectangular sensor network, while the inverse process is described in Section \ref{sec:K2G8SRect2Circ}.

\paragraph{Circular to Rectangular}\label{sec:K2G8SCirc2Rect}\hfill\\
%
%

In this case study, a dimensionality reduction of 99\% was achieved, with image dimensions reduced from $7 \times 10,568$ to $7 \times 897$.
Since the plate material is the same as in Section \ref{sec:G16Circ2Rect} and only the sensor network differs, similar results were anticipated.
This expectation is confirmed by the statistical metrics in Table \ref{tab:K2G8S_Circ_to_Rect_dist}, which assess the similarity between source and target domain distributions before and after MPCA.
As with the $G16$ plate in Section \ref{sec:G16Circ2Rect}, most distance measures increase following MPCA, indicating moderate distortion in the data distributions.
An exception is the Earth Mover’s Distance ($EMD$), which decreases, suggesting that MPCA improves the geometric alignment between distributions despite minor perturbations in their probability densities — as indicated by increases in metrics such as $KL$, $JSD$, and $\chi^2$.
Given the consistent improvements observed in Section \ref{sec:G16} when MPCA is combined with fine-tuning, a similarly high level of predictive performance can be expected in this scenario.

\begin{table}[hbtp]
    \centering
    \caption{Statistical metrics to assess the similarity between source (circular sensor network) and target (rectangular sensor network) domain data distributions.}
    \label{tab:K2G8S_Circ_to_Rect_dist}
    \begin{tabular}{l|cc}
        \begin{tabular}{c} Statistical \\ Distance\end{tabular} & No MPCA & 99\% MPCA \\
        \hline
        $KL$ & $7.45 \times 10^{-4}$ & $1.10 \times 10^{-2}$\\
        $KL_{sym}$ & $4.29 \times 10^{-4}$ & $1.02 \times 10^{-2}$\\
        $JSD$ & $3.23 \times 10^{-5}$ & $5.27 \times 10^{-4}$\\
        $\chi^2$ & $1.13 \times 10^{-4}$ & $1.57 \times 10^{-3}$\\
        $B$ & $3.86 \times 10^{-5}$ & $7.09 \times 10^{-4}$\\
        $EMD$ & $2.16 \times 10^{-5}$ & $8.26 \times 10^{-6}$\\
    \end{tabular}
\end{table}

\begin{figure}[htbp]
    \centering
    \includegraphics[width=\columnwidth]{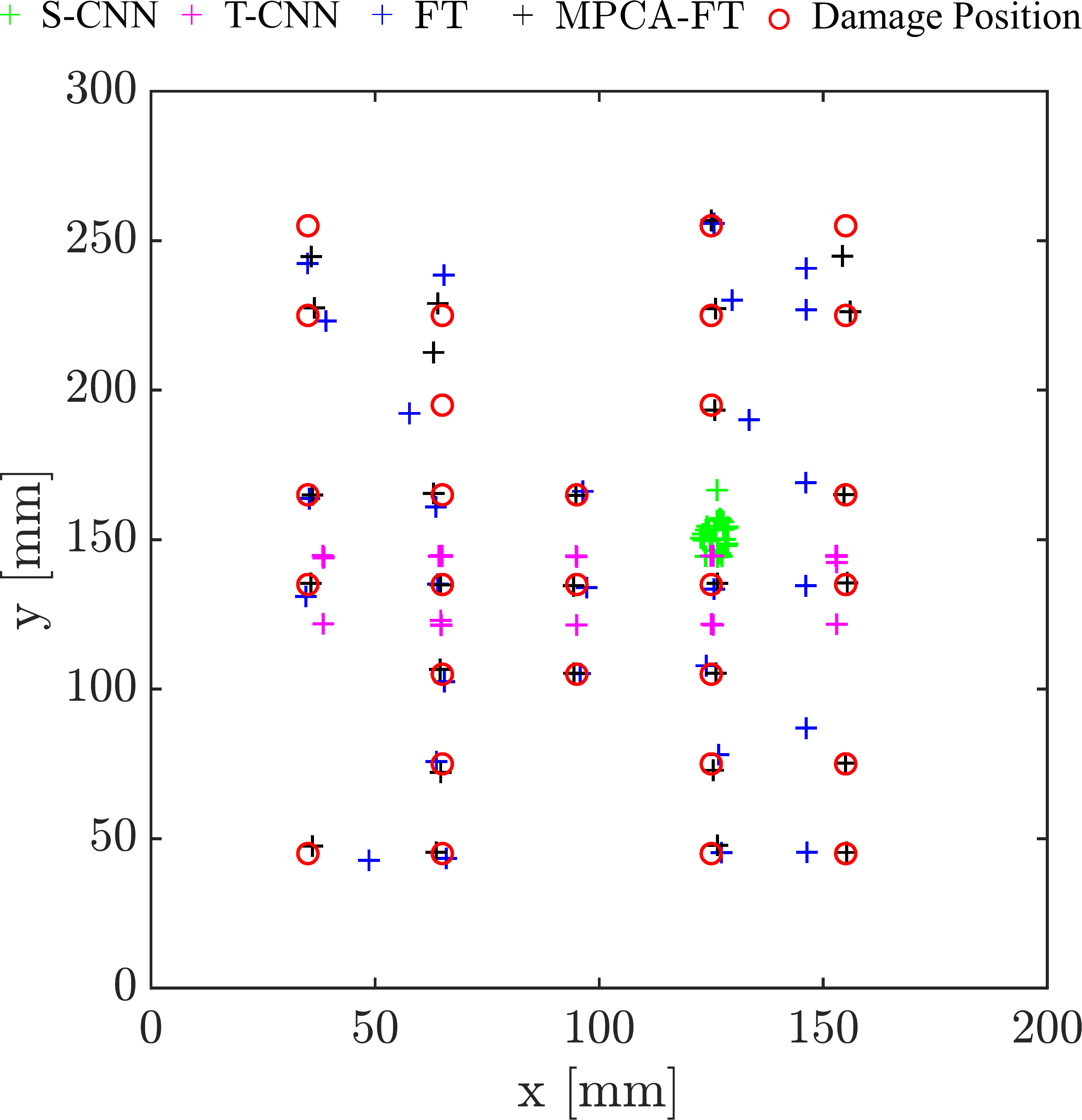}
    \caption{Damage positions predicted by all CNNs.}
    \label{fig:K2G8S_Circ_Rect}
\end{figure}

Turning to the analysis of localisation performance, Figure \ref{fig:K2G8S_Circ_Rect} shows the actual damage locations as red circles (\textcolor{red}{$\rm{\circ}$}), along with the predicted positions from (i) \textit{S-CNN} (\textcolor{green}{$\rm{+}$}), (ii) \textit{T-CNN} (\textcolor{magenta}{$\rm{+}$}), (iii) \textit{FT} (\textcolor{blue}{$\rm{+}$}), and (iv) \textit{MPCA-FT} (black crosses, $\rm{+}$).
The figure clearly illustrates that both \textit{S-CNN} and \textit{T-CNN} fail to provide reliable localisation, as their predictions are clustered in the central region of the plate and do not correspond to the true damage locations.
In contrast, fine-tuning significantly improves prediction accuracy, as reflected in the \textit{FT} outputs and confirmed by the error values in Table \ref{tab:KSG8S_circ_to_rect}.
While the \textit{FT} model yields satisfactory results, with prediction errors slightly above 5 mm in both directions, the combination of MPCA and fine-tuning (\textit{MPCA-FT}) achieves the lowest errors overall.
This is visually supported by Figure \ref{fig:K2G8S_Circ_Rect}, where the $\rm{+}$ predictions from \textit{MPCA-FT} are consistently closer to the true damage locations (\textcolor{red}{$\rm{\circ}$}) than those from \textit{FT} (\textcolor{blue}{$\rm{+}$}).
These findings provide strong evidence for the effectiveness of MPCA as a domain adaptation technique when used in conjunction with transfer learning.
The results demonstrate that the combination of MPCA and FT offers improved localisation capabilities compared to fine-tuning alone, even in scenarios where the sensor network changes but the material remains the same.

\begin{table}[htbp]
    \centering
    \caption{Root mean square localisation error along the $x$ and $y$ axes for each CNN configuration, evaluated on the target (rectangular sensor array) domain data. Lower values indicate better localisation performance.}
    \label{tab:KSG8S_circ_to_rect}
    \begin{tabular}{l|cc}
        & X Error [$\mathrm{mm}$] & Y Error [$\mathrm{mm}$] \\
        \hline
        S-CNN       & 52.14 & 65.51 \\
        T-CNN       & 1.75 & 60.79 \\
        FT           & 5.47 & 5.48  \\
        MPCA-FT    & 0.96 & 4.57  \\
    \end{tabular}
\end{table}

\paragraph{Rectangular to Circular}\label{sec:K2G8SRect2Circ}\hfill\\
%
%

The application of 99\% MPCA results yields a final image size of $7 \times 81$.
This level of compression is comparable to that observed for the $G16$ plate in Section \ref{sec:G16Rect2Circ}.
Following MPCA, all statistical distance metrics increase relative to their pre-MPCA values, indicating a distortion in the data distributions.
Specifically, the Earth Mover’s Distance ($EMD$) increases by one order of magnitude, while the other metrics (e.g., $KL$, $JSD$, and $\chi^2$) increase by a factor of two to three times greater than that of $EMD$.
This behaviour is consistent with the pattern observed for $G16$ in Section \ref{sec:G16Rect2Circ}, suggesting that the impact of MPCA on distributional similarity is similarly pronounced in both cases.

\begin{table}[hbtp]
    \centering
    \caption{Statistical metrics to assess the similarity between source (rectangular sensor network) and target (circular sensor network) domain data distributions.}
    \label{tab:K2G8S_Rect_to_Circ_dist}
    \begin{tabular}{l|cc}
        \begin{tabular}{c} Statistical \\ Distance\end{tabular} & No MPCA & 99\% MPCA \\
        \hline
        $KL$ & $1.13 \times 10^{-4}$ & $2.47 \times 10^{-1}$\\
        $KL_{sym}$ & $4.29 \times 10^{-4}$ & $1.80 \times 10^{-1}$\\
        $JSD$ & $3.25 \times 10^{-5}$ & $7.70 \times 10^{-3}$\\
        $\chi^2$ & $1.14 \times 10^{-4}$ & $2.36 \times 10^{-2}$\\
        $B$ & $3.88 \times 10^{-5}$ & $1.05 \times 10^{-2}$\\
        $EMD$ & $2.17 \times 10^{-5}$ & $3.05 \times 10^{-4}$\\
    \end{tabular}
\end{table}

\begin{figure}[htbp]
    \centering
    \includegraphics[width=\columnwidth]{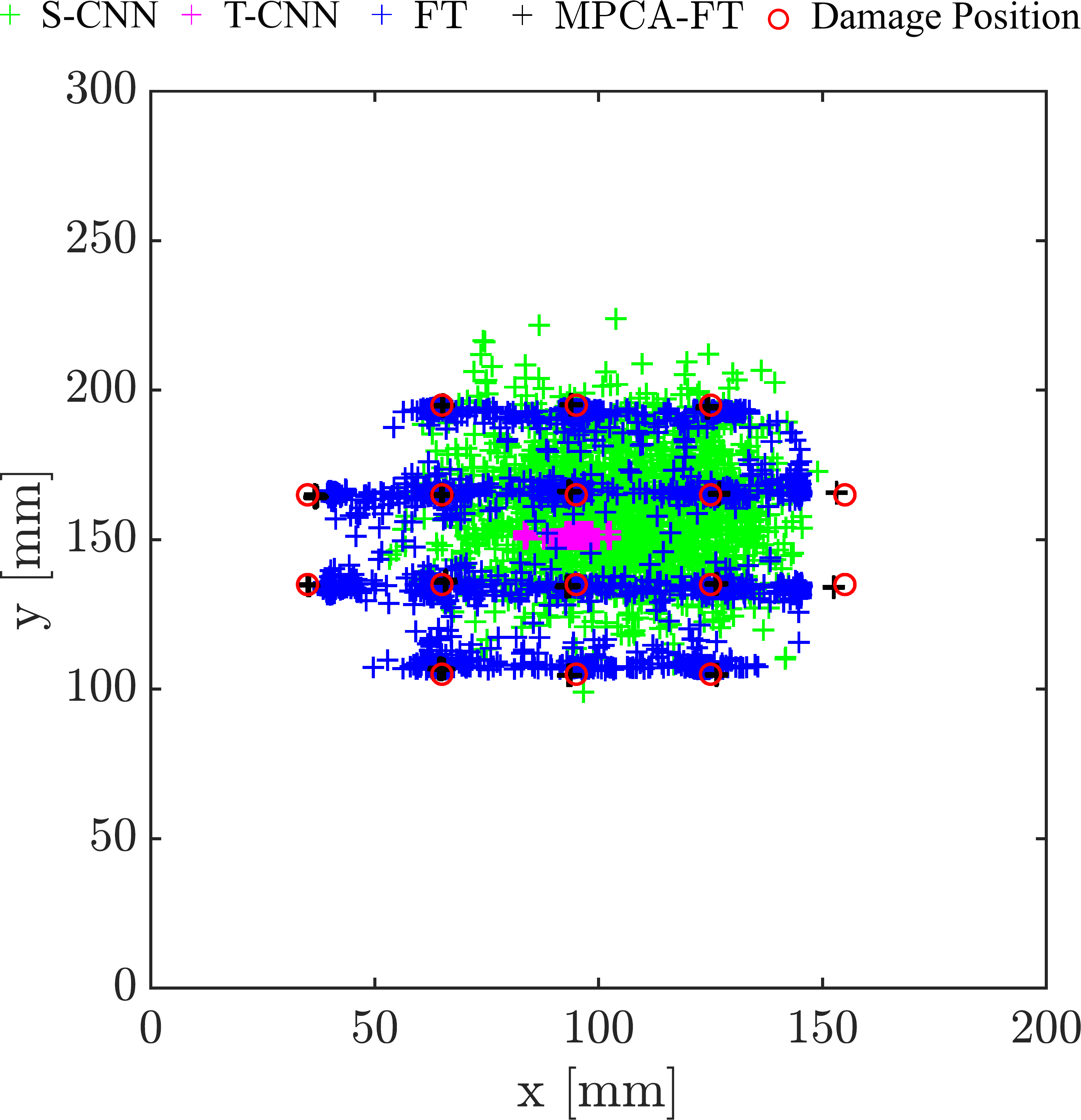}
    \caption{Damage positions predicted by all CNNs.}
    \label{fig:K2G8S_Rect_Circ}
\end{figure}

Given the comparable dimensionality reduction and distributional effects observed in both cases, it is reasonable to expect similar performance outcomes across these case studies.
As anticipated, Figure \ref{fig:K2G8S_Rect_Circ} replicates the behaviour observed in Figure \ref{fig:G16_Rect_Circ} for the $G16$ plate.
The predictions produced by \textit{S-CNN}, shown as \textcolor{green}{$\rm{+}$}, are randomly scattered, indicating an inability to localise damage reliably.
In contrast, \textit{T-CNN} predictions (\textcolor{magenta}{$\rm{+}$}) are concentrated near the centre of the plate, but still fail to align with the actual damage positions.
The poor performance of both models is further supported by the large localisation errors reported in Table \ref{tab:KSG8S_rect_to_circ} for both $x$ and $y$ coordinates.
Fine-tuning (\textit{FT}) achieves comparatively lower error values; however, the spatial distribution of its predictions (\textcolor{blue}{$\rm{+}$}) in Figure \ref{fig:K2G8S_Rect_Circ} reveals that these lower errors are not due to accurate localisation.
Instead, they result from the random distribution of predicted $x$-coordinates across the correct $y$-positions, which artificially reduces error magnitude.
By contrast, \textit{MPCA-FT} delivers the most accurate localisation performance, achieving prediction errors of 1.39 $mm$ and 0.73 $mm$ in the $x$ and $y$ directions, respectively.
This is visually confirmed in Figure \ref{fig:K2G8S_Rect_Circ}, where the black crosses ($\rm{+}$) are consistently positioned close to the actual damage locations (\textcolor{red}{$\rm{\circ}$}).
These findings further confirm that the combination of MPCA and fine-tuning provides the most effective solution for transfer learning (TL) of UGW data across different materials with the same sensor array and different sensor networks for the same material.
This result underscores the potential of MPCA not only as a dimensionality reduction tool but also as a powerful component in domain adaptation and parameter-based TL pipelines.

\begin{table}[htbp]
    \centering
    \caption{Root mean square localisation error along the $x$ and $y$ axes for each CNN configuration, evaluated on the target (circular sensor array) domain data. Lower values indicate better localisation performance.}
    \label{tab:KSG8S_rect_to_circ}
    \begin{tabular}{l|cc}
        & X Error [$\mathrm{mm}$] & Y Error [$\mathrm{mm}$] \\
        \hline
        S-CNN       & 42.42 & 38.90 \\
        T-CNN       & 29.13 & 29.06 \\
        FT           & 13.13 & 4.33  \\
        MPCA-FT    & 1.39 & 0.73  \\
    \end{tabular}
\end{table}

\section{Conclusions}\label{sec:conc}
This work investigated the applicability of Multilinear Principal Component Analysis (MPCA) for transfer learning (TL) in the context of data scarcity.
The novelty of the proposed approach lies in leveraging a method traditionally used for dimensionality reduction to perform domain adaptation (DA) and dimensionality reduction simultaneously.
To assess the potential of MPCA for TL, a dataset of Ultrasonic Guided Wave (UGW) signals was collected from three composite plates made of different materials ($G16$, $K8$, and $K2G4S$). Each plate was instrumented with both circular and rectangular sensor networks.
In the experiments, the source domain consisted of all available data for a given material–sensor configuration, while the target domain included only half of the available data.
Statistical metrics were employed to quantify the similarity between source and target domain data distributions before and after MPCA application.
Damage localisation performance was assessed by computing the Root Mean Square Error (RMSE) between the predicted and actual damage positions along both spatial coordinates.
The key findings of this study are as follows:
\begin{itemize}

    \item MPCA can be effectively employed as a domain adaptation technique.
    The combination of MPCA and fine-tuning (FT) consistently outperformed FT alone, suggesting that MPCA helps mitigate inter-domain differences.
    
    \item The statistical metrics used to evaluate distributional similarity may not fully capture the positive impact of MPCA.
    In several cases, statistical distances increased after MPCA, despite clear improvements in localisation performance.
    
    \item Performing TL between $G16$ and $K8$, or vice-versa, implies a lower dimensionality reduction than moving from either $G16$ or $K8$ to $K2G4S$ and vice-versa.
    This means that the use of a hybrid material ($K2G4S$) increases feature overlap, as could be expected. 
    
\end{itemize}
In summary, this study demonstrates the feasibility and effectiveness of MPCA as a domain adaptation technique for TL in UGW-based Structural Health Monitoring.
Future work may explore comparisons with other DA techniques and extend the analysis to additional case studies, including both regression and classification tasks.

\section*{Declaration of Competing Interest}
The authors declare that they have no known competing financial interests or personal relationships that could have appeared to influence the work reported in this paper.

\section*{Data Availability Statement}
The datasets generated and analysed during the current study are available in the Zenodo repository named UGW-3Mat-2SN \cite{UGW3MAT2SN}. They can be accessed via the DOI: \url{https://doi.org/10.5281/zenodo.15688321}. 


\end{document}